\documentclass[superscriptaddress,prl,floatfix,twocolumn,amsmath,amssymb,aps]{revtex4-1}

\usepackage{graphicx}
\usepackage{amsmath}
\usepackage{amssymb}
\usepackage{dcolumn}
\usepackage{hyperref}
\usepackage{color}
\usepackage{multirow}
\usepackage{bm}
\usepackage{subfigure}
\usepackage[normalem]{ulem}

\begin{document}
\title{Crystal-Structure Matches in Solid-Solid Phase Transitions}
\author{Fang-Cheng Wang}
\affiliation{State Key Laboratory for Artificial Microstructure and Mesoscopic Physics, Frontier Science Center for Nano-optoelectronics, School of Physics, Peking University, Beijing 100871, People's Republic of China}
\author{Qi-Jun Ye}
\email{qjye@pku.edu.cn}
\affiliation{State Key Laboratory for Artificial Microstructure and Mesoscopic Physics, Frontier Science Center for Nano-optoelectronics, School of Physics, Peking University, Beijing 100871, People's Republic of China}
\affiliation{Interdisciplinary Institute of Light-Element Quantum Materials, Research Center for Light-Element Advanced Materials, and Collaborative Innovation Center of Quantum Matter, Peking University, Beijing 100871, People's Republic of China}
\author{Yu-Cheng Zhu}
\affiliation{State Key Laboratory for Artificial Microstructure and Mesoscopic Physics, Frontier Science Center for Nano-optoelectronics, School of Physics, Peking University, Beijing 100871, People's Republic of China}
\author{Xin-Zheng Li}
\email{xzli@pku.edu.cn}
\affiliation{State Key Laboratory for Artificial Microstructure and Mesoscopic Physics, Frontier Science Center for Nano-optoelectronics, School of Physics, Peking University, Beijing 100871, People's Republic of China}
\affiliation{Interdisciplinary Institute of Light-Element Quantum Materials, Research Center for Light-Element Advanced Materials, and Collaborative Innovation Center of Quantum Matter, Peking University, Beijing 100871, People's Republic of China}
\affiliation{Peking University Yangtze Delta Institute of Optoelectronics, Nantong, Jiangsu 226010, People's Republic of China}
\date{\today}

\begin{abstract}
    The exploration of solid-solid phase transition suffers from the uncertainty of how atoms in two crystal structures match.
    We devised a theoretical framework to describe and classify crystal-structure matches (CSM).
    Such description fully exploits the translational and rotational symmetries and is independent of the choice of supercells.
    This is enabled by the use of the Hermite normal form, an analog of reduced echelon form for integer matrices.
    With its help, exhausting all CSMs is made possible, which goes beyond the conventional optimization schemes.
    In an example study of the martensitic transformation of steel, our enumeration algorithm finds many candidate CSMs with lower strains than known mechanisms.
    Two long-sought CSMs accounting for the most commonly observed Kurdjumov-Sachs orientation relationship and the Nishiyama-Wassermann orientation relationship are unveiled.
    Given the comprehensiveness and efficiency, our enumeration scheme provide a promising strategy for solid-solid phase transition mechanism research.
\end{abstract}

\maketitle
\begin{figure}[t]
    \centering
    \includegraphics[width=\linewidth]{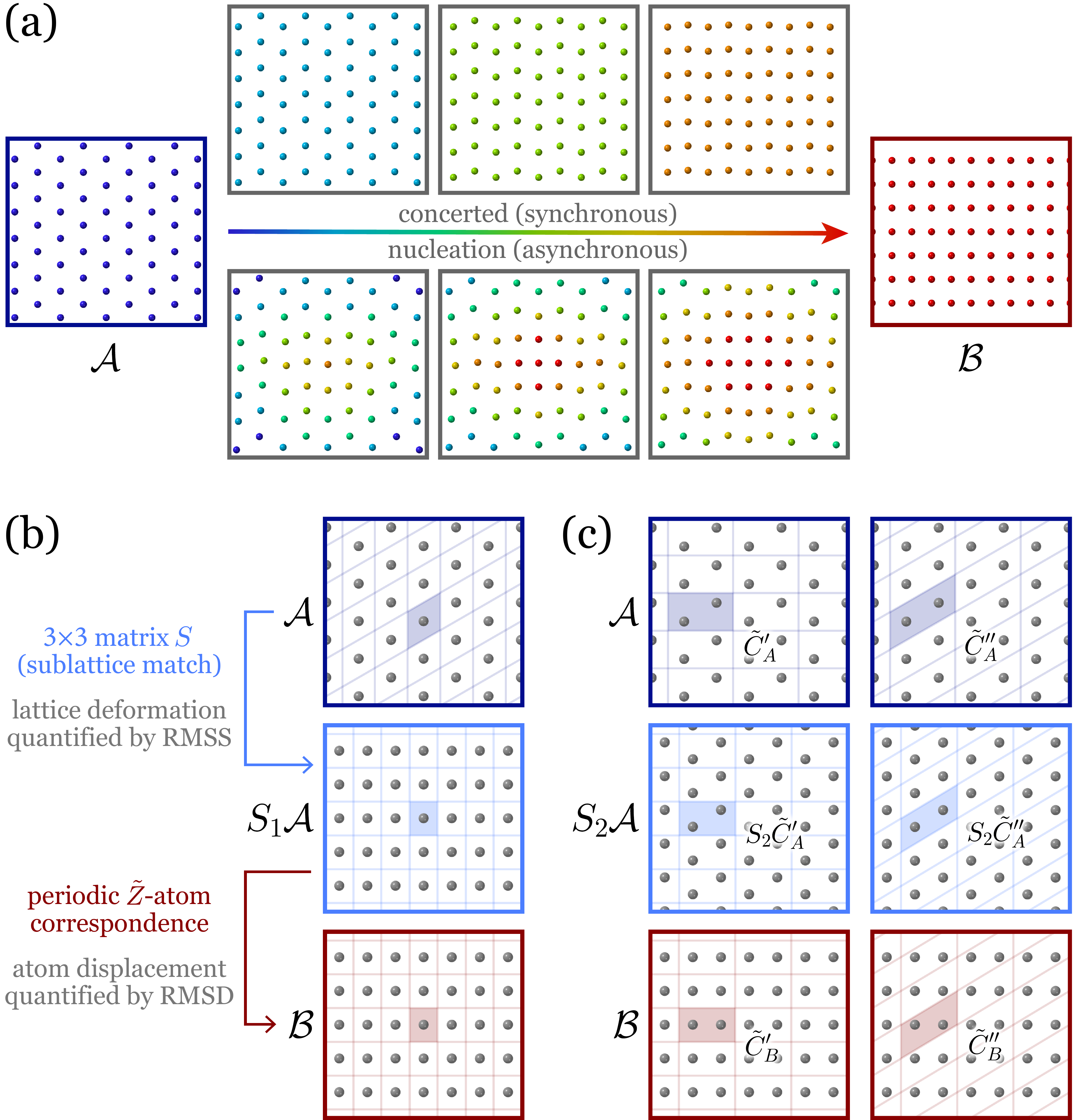}
    \caption{CSMs between $\mathcal{A}$ (hexagonal) and $\mathcal{B}$ (orthogonal). (a) The same CSM can be established either by concerted mechanisms or by nucleation, where the atoms migrate to their counterparts synchronously or asynchronously. The colors of the atoms represent their own degrees of migration. (b) A CSM consists of an SLM which matches the sublattice of $\mathcal{A}$ to that of $\mathcal{B}$, and a periodic $\tilde Z$-atom correspondence. In this example we have $\tilde Z=1$. (c) Another CSM with $\tilde Z=2$, whose RMSS is lower but RMSD is higher than (b). The two columns show different supercell pair choices, respectively.}
    \label{Fig1}
\end{figure}

Solid-solid phase transition (SSPT) is ubiquitous in nature and relevant to many industries~\cite{callister2000fundamentals}.
For example, the martensitic transition is a process of immense importance in the steel industry~\cite{greninger1949mechanism,nishiyama2012martensitic,therrien2020minimization}, and the graphite-to-diamond transition under shock compression enables the synthesis of highly desirable diamond from abundant and cheap carbon sources~\cite{decarli1961formation,erskine1991shock,scandolo1995pressure,mundy2008ultrafast}.
Compared to other well-studied dynamical processes, such as the gas-phase or surface reactions~\cite{eyring1935activated,jonsson1998nudged,henkelman2000climbing,heyden2005efficient}, SSPTs involve not only much greater degrees of freedom (d.o.f.)~\cite{khaliullin2011nucleation,badin2021nucleating,santos2022size} but also complex collective behavior with controversial mechanisms~\cite{therrien2021metastable}.
Besides, the crystalline nature highlights the significance of lattice deformation in the reaction path, which is distinct from the fluid-solid or fluid-fluid phase transitions~\cite{sheppard2012generalized,poole1997polymorphic}.
With potential insights into the mechanisms and principles of self-organization, symmetry breaking, and criticality in multiple disciplines, understanding the atomic details of SSPTs is urgent but remains in its infancy.
Theoretical studies of SSPTs fall into two categories: nucleation and concerted mechanisms.
Although a realistic SSPT generally occurs through nucleation, the vast d.o.f. it involves are expensive for atomic-level simulations.
The concerted mechanism is a simplified model with few d.o.f. in a small supercell with periodic boundary condition, and thus can be investigated at the density-functional theory level~\cite{sheppard2012generalized}.
Despite their differences, simulations using ${\sim}10^5$ atoms suggest that minimum energy paths (MEP) of nucleation proceed locally via much simpler concerted mechanisms~\cite{khaliullin2011nucleation,badin2021nucleating}.
That is to say, nucleation is likely to share the same atom-to-atom correspondence between the initial and final structures---which we call the crystal-structure match (CSM)---with a concerted mechanism [see Fig.~\ref{Fig1}(a)].
As a universal concept, the CSM lies at the heart of the SSPT research.
This is especially true for concerted mechanisms, where methods to find the MEP like the solid-solid nudged elastic band (SSNEB) require the user to prespecify a pair of supercells in two phases and the correspondence between the atoms in them~\cite{sheppard2012generalized,zhang2015variable}.
However, human intuition often fails to select the ``best'' CSM, in which case the obtained MEP is not the global one.
To address this issue, methods like the stochastic surface walking~\cite{guan2015energy} and {\sc pallas}~\cite{zhu2019phase} try different CSMs as they explore the potential energy surface.
Meanwhile, several studies have been dedicated to design criteria for instructing CSM without explicit energy minimization, e.g., maximal symmetry~\cite{capillas2007maximal}, minimal strain~\cite{chen2016determination}, minimal dissociation of chemical bonds~\cite{stevanovic2018predicting}, and minimal total distance traveled by the atoms~\cite{therrien2020matching}.
Nevertheless, existing optimization algorithms cannot claim conclusively to find the best CSMs under these criteria, nor are they guaranteed to be energetically favorable.

In this Letter, we provide a theoretical framework to describe, classify, and enumerate CSMs.
We call the lattice deformation of a CSM its sublattice match (SLM), since it deforms certain sublattices of the initial crystal structure into sublattices of the final one [see Fig.~\ref{Fig1}(b)].
By extracting key features of a sublattice as a Hermite normal form (HNF)~\cite{cohen2013course,SM}, we show that every SLM can be uniquely represented by an integer-matrix triplet (IMT).
This eliminates the heavy redundancy in the conventional supercell pair representation~\cite{sheppard2012generalized,chen2016determination,li2022smallest}, making it possible to \textit{exhaust} all SLMs within a certain range.
For each SLM, its representative CSM is obtained via the Hungarian algorithm which minimizes the atomic displacement~\cite{kuhn1955hungarian}.
Using this strategy, we provide a comprehensive list of CSMs in an example study of the martensitic transformation of steel.
Among the enumerated CSMs, we discover the ones that account for the most commonly observed Kurdjumov-Sachs (KS) orientation relationship (OR)~\cite{kurdjumow1930mechanismus} and the Nishiyama-Wassermann (NW) OR~\cite{nishiyama2012martensitic}, as well as ones with much lower strains than all previously known mechanisms.
Practically, we consider a CSM with the following property~\footnote{Otherwise, the CSM must be described in an infinitely large supercell, which has $\tilde Z=\infty$ and is out of the reach of simulations using finitely many atoms. Concerted mechanisms always have finite $\tilde Z$'s, while $\tilde Z\le 12$ generally holds within simulated critical nuclei~\cite{khaliullin2011nucleation,badin2021nucleating,santos2022size}.}: there exists a linear transformation $S$ such that after its deformation, the atoms in the deformed structure $S\mathcal{A}$ correspond to the atoms in $\mathcal{B}$ \textit{periodically}, as shown in Figs.~\ref{Fig1}(b) and \ref{Fig1}(c).
Denote the number of atoms in the smallest spatial period by $\tilde Z$.
As $S$ transforms the sublattices in $\mathcal{A}$ into those in $\mathcal{B}$, there exist pairs of $\tilde Z$-atom supercells such that
\begin{equation}\label{scacb}
  S\tilde C_A=\tilde C_B
,\end{equation}
where $\tilde C_\alpha\pod{\alpha=A\text{ or }B}$ is a $3\times 3$ matrix whose columns are supercell vectors arranged in right-handed order.
We call $S$---with the sublattices before and after deformation---the sublattice match (SLM) of this CSM.
Denoting a primitive cell in $\alpha$ by $C_\alpha$, the supercell is associated with an integer matrix $M_\alpha$ as $\tilde C_\alpha=C_\alpha M_\alpha$~\footnote{A supercell vector is always an integer linear combination of primitive-cell vectors, as $\tilde{\mathbf{v}}_{i} = \sum_{j=1}^3 m_{ji} \mathbf{v}_{j}$, where $m_{ij}\in \mathbb{Z}$ are elements of $M_\alpha$. These vectors form the columns of $\tilde C_\alpha$ and $C_\alpha$ in right-handed order, making the determinants of $\tilde C_\alpha$, $C_\alpha$ and $M_\alpha$ positive.} with the constraint
\begin{equation}\label{zcon}
  Z_\alpha\det M_\alpha=\tilde Z
,\end{equation}
where $Z_\alpha$ is the number of atoms in $C_\alpha$.
This means that each $(M_A,M_B)$ satisfying Eq.~(\ref{zcon}) gives an SLM, as
\begin{equation}\label{scp}
  S=C_BM_BM_A^{-1}C_A^{-1}
,\end{equation}
which we call the supercell pair representation~\cite{sheppard2012generalized,chen2016determination,li2022smallest}.
However, such supercell pair of a given SLM is not unique [see Fig.~\ref{Fig1}(c)].
This is because for any integer matrix $Q$ with $\det Q=1$, another $(M_A',M_B')=(M_AQ,M_BQ)$ also gives the same $S$ and satisfies Eq.~(\ref{zcon}).
This redundancy has made the enumeration of SLMs extremely difficult, as reported in Ref.~\cite{li2022smallest}.
We overcome this difficulty by utilizing the theorem: any integer matrix $M$ with $\det M>0$ can be \textit{uniquely} decomposed into two integer matrices as $M=HQ$, where $H$ is in Hermite normal form (HNF) and $\det Q=1$~\cite{cohen2013course}.
That is to say, under elementary column operations over integers~\cite{SM}, $M$ can be uniquely transformed into
\begin{equation}\label{hnf}
  H=\begin{bmatrix}
	  h_{11}&0&0\\h_{21}&h_{22}&0\\h_{31}&h_{32}&h_{33}
  \end{bmatrix},\quad 0\le h_{ij}<h_{ii}\pod{j<i}
,\end{equation}
which is an analog of reduced echelon form for integer matrices.
Similar techniques have been used in the Hart-Forcade theory, a well-developed framework for generating derivative structures~\cite{santoroCoincidencesiteLattices1973,santoroPropertiesCrystalLattices1972,hart2008algorithm}.
Applying $M_\alpha=H_\alpha Q_\alpha$ to Eq.~(\ref{scp}), we obtain
\begin{equation}\label{imt}
  S=C_BH_BQH_A^{-1}C_A^{-1}
,\end{equation}
where $Q=Q_BQ_A^{-1}$ is an integer matrix with $\det Q=1$ and $H_\alpha$ is in HNF satisfying
\begin{equation}\label{deth}
	\det H_\alpha = \frac{\tilde Z}{Z_\alpha}
,\end{equation}
which is derived from Eq.~(\ref{zcon}).
This IMT representation $(H_A,H_B,Q)$ of a given SLM is \textit{unique}, with which one can prove that the total number of $\tilde Z$-atom SLMs is finite and thus exhaustible as long as the strain is bounded.
For mathematical details, see our Supplemental Material~\cite{SM}.
To quantify the strain, we use the quadratic average of the principal strains (PS) of $S$, which we call the root-mean-square strain (RMSS), but one could well use other criteria.
Given an SLM, the only unspecified part of a CSM is how the $\tilde Z$ atoms in $\tilde C_A$ are mapped to atoms in $\mathcal{B}$~\footnote{Regardless of the choice of supercell, specifying the counterparts of $\tilde Z$ inequivalent atoms always determines a CSM. Therefore, for a given SLM $(H_A,H_B,Q)$, one may simply let $\tilde C_A=C_AH_A$.}.
Note that the sublattice given by the SLM divides atoms in $\mathcal{B}$ into $\tilde Z$ translational equivalence classes.
There are at most $\tilde Z!$ possible ways for a CSM to assign $\tilde Z$ atoms in $\tilde C_A$ to the $\tilde Z$ equivalence classes in $\mathcal{B}$.
When it is unfeasible to check numerous correspondences one by one, there is often a focus on the best one under certain geometric criterion, e.g., the root-mean-square displacement (RMSD) from the atoms in $S\mathcal{A}$ to their counterparts in $\mathcal{B}$~\cite{SM}.
We call such best CSM the representative of its SLM, which can always be solved via the Hungarian algorithm in polynomial time~\cite{kuhn1955hungarian}.
We use the RMSS and RMSD to characterize CSMs because they are proportional to the Euclidean distance between structures contributed by the lattice d.o.f. and atomic d.o.f., respectively~\cite{SM}.

\begin{figure}[b]
    \centering
    \includegraphics[width=\linewidth]{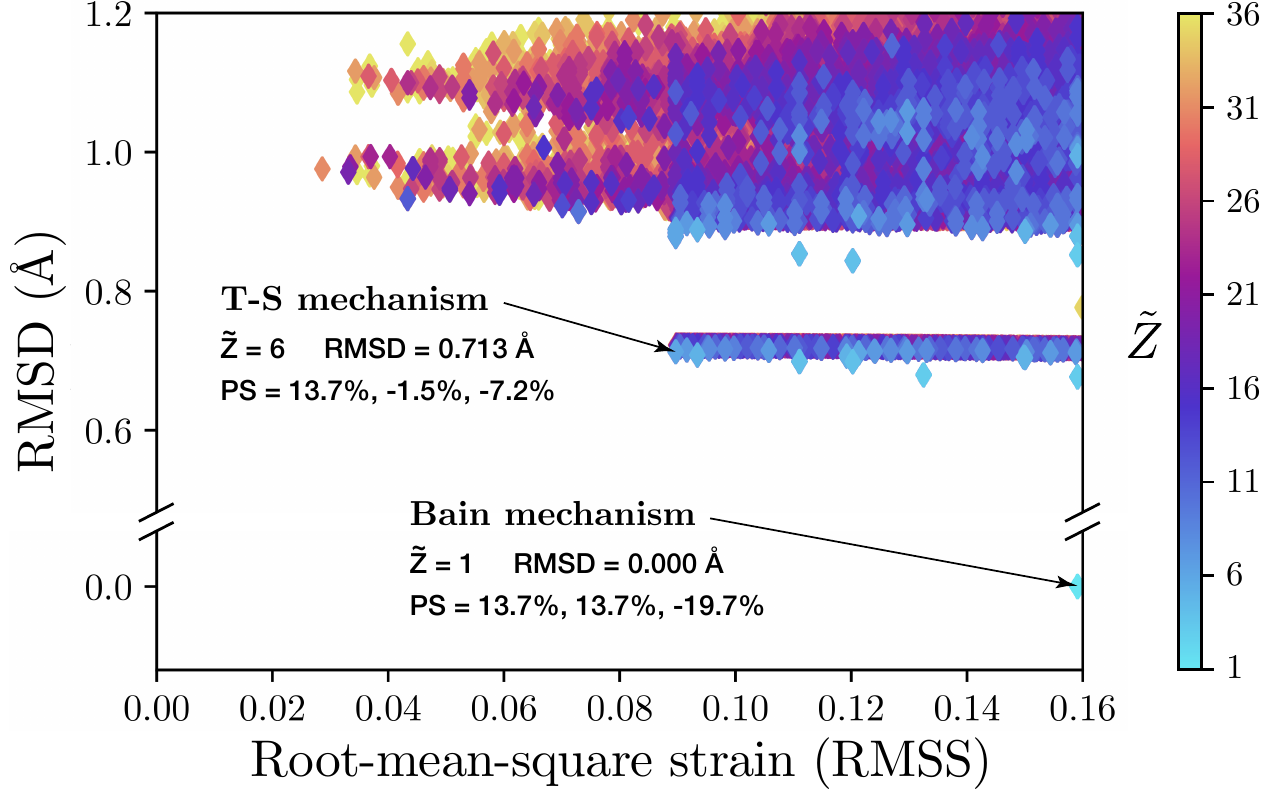}
    \caption{CSMs of the martensitic transformation of steel. For simplicity, only the representative CSM of each SLM is shown. Arrows point out the Bain mechanism~\cite{bain1924nature} and the Therrien-Stevanovi{\'c} one~\cite{therrien2020minimization}. Besides them, the vast remaining CSMs have not been reported previously, which might imply new mechanisms.}
    \label{Fig2}
\end{figure}

In an example study of the martensitic transformation of steel, an SSPT from the austenite phase (fcc, $a_\text{fcc}=3.57~\text\AA$) to the martensite phase (bcc, $a_\text{bcc}=2.87~\text\AA$), we exhaust all SLMs with $\tilde Z\le 36$ and $\mathrm{RMSS}\le 16\%$~\cite{SM}.
With 54\,178 SLMs enumerated, we compute the representative CSM for each SLM, as shown in Fig.~\ref{Fig2}.
The exhaustion of SLMs is accomplished through the IMT representation:
(i) Randomly generate a trial linear transformation $S_0$ with $\mathrm{RMSS}\le 16\%$.
(ii) Exhaust all $H_\alpha$'s satisfying Eqs.~(\ref{hnf}) and (\ref{deth}). For each $(H_A,H_B)$, compute the nearest integer matrix to the solution of Eq.~(\ref{imt}), namely
\begin{equation}
	Q_0=\operatorname{rint}(H_B^{-1}C_B^{-1}S_0C_AH_A)
,\end{equation}
where ``rint'' rounds each matrix element to its nearest integer. 
If $\det Q_0=1$ holds, an SLM represented by $(H_A,H_B,Q_0)$ is obtained.
(iii) Repeat this process until the number of consecutive replications of SLMs reaches the convergence criterion.
The implementation is available as a Python package {\sc crystmatch}~\cite{code}.
To reveal the realistic CSMs, we compare these enumeration results to experiments via orientation relationship (OR) analysis.
The OR specifies how the crystallographic axes of the product phase are oriented relative to the parent phase.
For experiments on the martensitic transformation of steel, the KS OR and/or NW OR are dominant in bulks~\cite{kurdjumow1930mechanismus,nishiyama2012martensitic,greninger1949mechanism}, while the Pitsch OR is only reported in thin films~\cite{pitsch1959martensite,therrien2020minimization}.
These ORs are denoted by parallelisms, as detailed in Ref.~\cite{koumatos2017theoretical}.
On the other hand, each CSM \textit{a priori} determines an OR through certain regulations.
Two most popular and reasonable manners are (i) keeping rotation-free and deforming $\mathcal{A}$ by $\sqrt{S^\text{T}S}$, which minimizes the total atomic displacement~\cite{SM}, as used in Ref.~\cite{therrien2020minimization}; (ii) imposing a rotation to restore the uniformly scaled plane (USP), as suggested by the phenomenological theory of martensitic transformation~\cite{khachaturyan2013theory}.
We shall show results under both manners.

\begin{figure}[b]
    \centering
    \includegraphics[width=\linewidth]{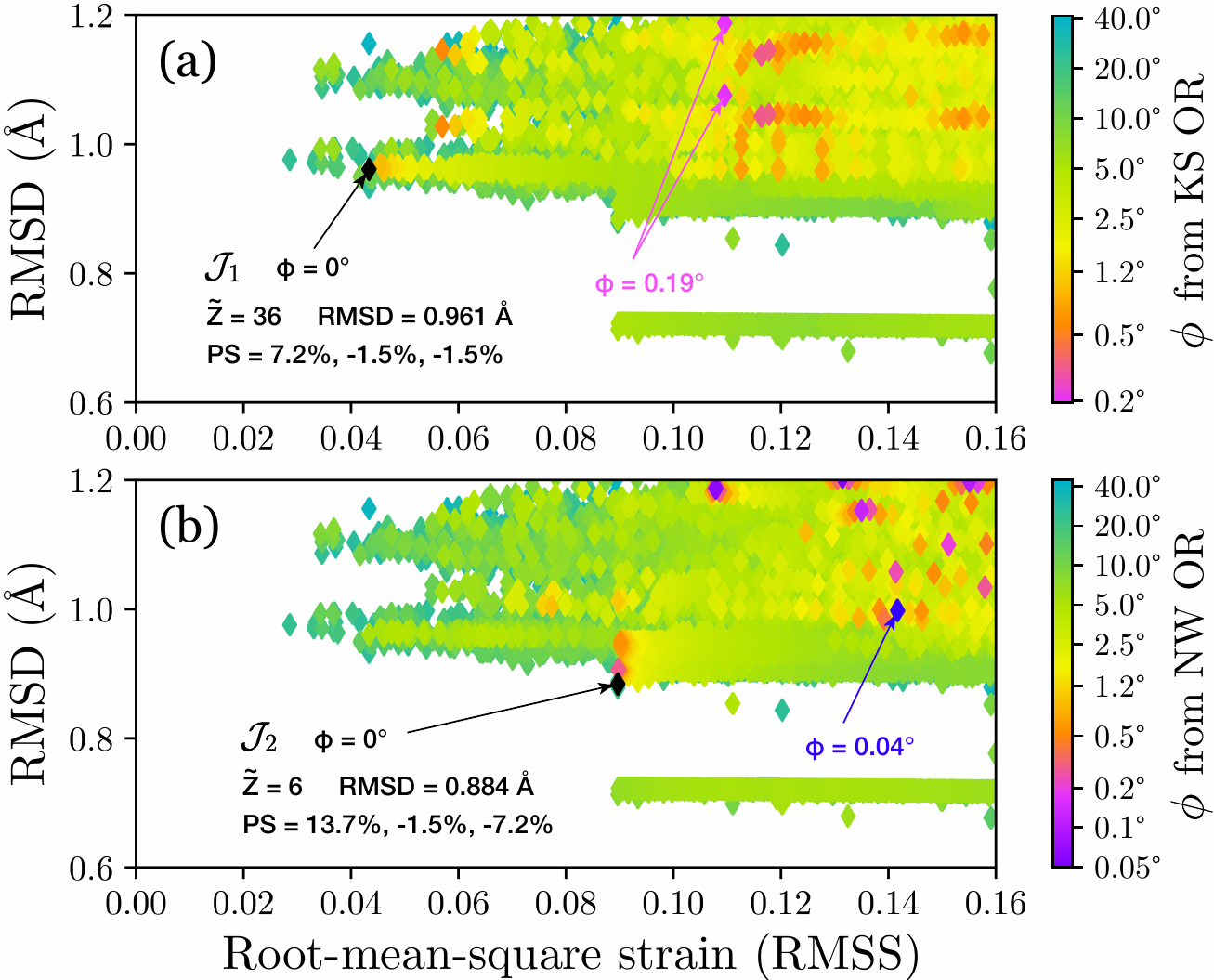}
    \caption{CSMs benchmarked by the (a) KS and (b) NW ORs. The distinction from an OR is measured by $\phi$, the least rotation angle required to produce that OR, as plotted with color bars. For either OR, we find only \textit{one} SLM with exact $\phi=0$, whose representative CSM is plotted in black as $\mathcal{J}_1$ and $\mathcal{J}_2$. Arrows point out all CSMs with $\phi$'s beyond the color bar. For reference, the average deviation of present OR measurements is $0.15^\circ$\cite{he2006observations}.}
    \label{Fig3}
\end{figure}

\begin{figure*}[t]
    \centering
    \includegraphics[width=0.9\linewidth]{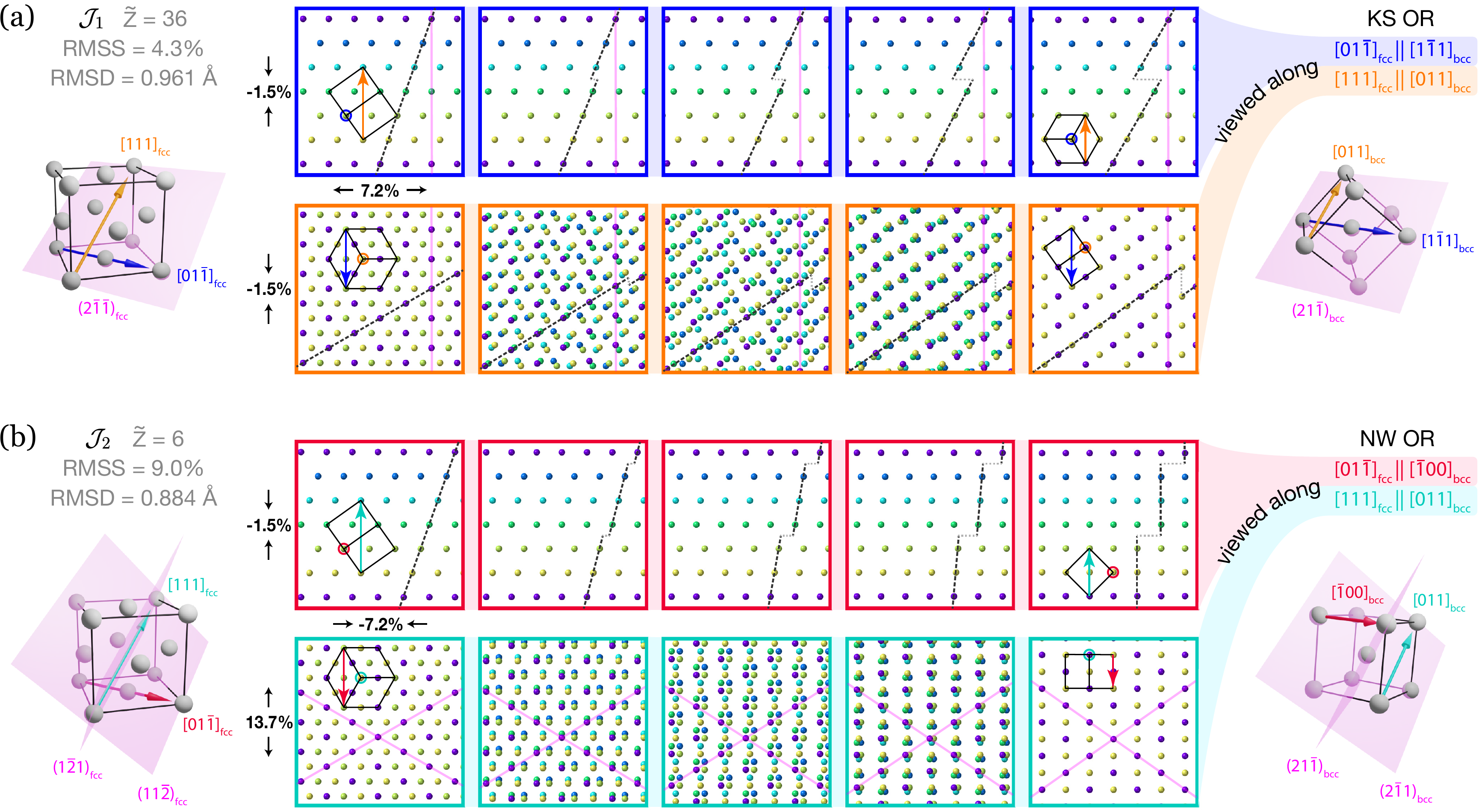}
    \caption{Concerted paths interpolated, respectively, from $\mathcal{J}_1$ and from $\mathcal{J}_2$.
	    The lattice deformation is assumed to be rotation-free as $\sqrt{S^\text{T}S}$, whose principal strains are labeled on the left side of the boxes.
	The colors of the atoms are used to distinguish different $(111)_\text{fcc}$ layers, and dashed lines are used to track the slipping process.
        Conventional cells of fcc and bcc are edged by solid black lines.
	USPs are denoted by pink planes.
	(a) $\mathcal{J}_1$ accounting for the KS OR, i.e., the parallelism $(111)[01\bar 1]_\text{fcc}\parallel(011)[1\bar 11]_\text{bcc}$.
(b) $\mathcal{J}_2$ accounting for the NW OR, i.e., the parallelism $(111)[01\bar 1]_\text{fcc}\parallel(011)[\bar 100]_\text{bcc}$.}
    \label{Fig4}
\end{figure*}

\begin{table}[b]
\caption{\label{tab:table1}%
The main features of the highlighted CSMs
}
\begin{ruledtabular}
\begin{tabular}{lccccc}
CSM&
$\tilde Z$&
RMSS&
RMSD~(\AA)&
OR\footnote{Using the rotation-free manner.}&
OR\footnote{Using the USP-restoring manner.}\\
\colrule
Bain & 1 & 15.9\% & 0 & Unreported & Unreported\\
T-S & 6 & 9.0\% & 0.713 & Pitsch & KS\\
$\mathcal{J}_1$ & 36 & 4.3\% & 0.961 & KS & KS\\
$\mathcal{J}_2$ & 6 & 9.0\% & 0.884 & NW & KS\\
\end{tabular}
\end{ruledtabular}
\end{table}

We benchmarked all CSMs by the KS, NW and Pitsch ORs, respectively.
Their distinctions from each OR are quantified by extra rotation angles, which are alone determined by their SLMs~\cite{therrien2020minimization,SM}.
Therefore, we only discuss the representative CSM of each SLM for simplicity.
When using the rotation-free manner, there is only one CSM that precisely conforms to each OR, as shown in Figs.~\ref{Fig3} and S4~\cite{SM}.
The CSM reproducing the Pitsch OR is consistent with the Therrien-Stevanovi{\'c} (T-S) mechanism~\cite{therrien2020minimization}, which validates our methods.
Besides, we found two new CSMs that reproduce the KS and NW ORs, which are denoted by $\mathcal{J}_1$ and $\mathcal{J}_2$ and will be discussed later.
When using another manner to restore the USP, we found that 55 CSMs can reproduce the KS OR, including $\mathcal{J}_1$, $\mathcal{J}_2$ and the T-S mechanism (see Fig.~S5~\cite{SM}), but no enumerated CSM can reproduce the NW or Pitsch OR.
The well-studied Bain mechanism~\cite{bain1924nature}, which has been proved to have the lowest RMSS among all $\tilde Z=1$ CSMs~\cite{koumatos2016optimality}, does not reproduce any observed OR in either manner.
Table~\ref{tab:table1} highlights the features of these CSMs, where the RMSS and RMSD are calculated from experimental lattice constants.
We note that the CSMs accounting for reported ORs have neither the lowest RMSS nor the lowest RMSD among all CSMs (see Figs.~\ref{Fig2} and \ref{Fig3}).
They are thus beyond the reach of conventional optimization schemes, which further demonstrates the necessity of enumeration.
To gain insight into the mechanisms therein, we illustrate the concerted paths interpolated, respectively, from $\mathcal{J}_1$ and from $\mathcal{J}_2$ in Fig.~\ref{Fig4}.
Both paths involve slipping processes along the $(111)_\text{fcc}\parallel(011)_\text{bcc}$ plane, which occur at every sixth layer and every third layer, respectively.
The $\mathcal{J}_1$ path has an additional intralayer slipping along the $[01\bar 1]_\text{fcc}\parallel[1\bar 11]_\text{bcc}$ direction.
The USPs of both $\mathcal{J}_1$ and $\mathcal{J}_2$ coincide with the experimentally observed $\{112\}_\text{fcc}$ habit plane~\cite{nishiyama2012martensitic,klostermann1964surface}.
Surprisingly, the USP in Fig.~\ref{Fig4}(a) stays still \textit{without} rotation, and even becomes a strictly invariant plane if close-packing lattice constants ($a_\text{bcc}=\sqrt{2 /3}\,a_\text{fcc}$) are used, which can explain the predominance of the KS OR.
Animations of both paths are provided in Supplemental Material~\cite{SM}, where we also apply our scheme to the wurtzite-zinc blende (B4-B3) transition in ZnS, a prototype SSPT involving multiple types of atoms.
Compared to yielding a single optimal CSM by conventional methods, an entire enumeration including the same CSM even takes less time in most cases.
Specifically, it takes only 29.17 \textit{seconds} on a single core (3.60 GHz) of Intel Core i7-12700K Desktop Processor to produce all CSMs with $\tilde Z\le 6$ in Fig.~\ref{Fig2}, which covers all previously reported CSMs.
We note that underlying this efficiency is the utilization of symmetries.
The translational symmetry is implied in the definition of SLM, where a sublattice is a subgroup of the translation group.
The exploitation of this crystalline feature frees us from the inherent difficulties in optimizing the rotation, deformation, translation and atomic correspondence simultaneously, and allows us to deal with $3\times 3$ matrices rather than manipulating large structures with many atoms.
By virtue of the rotational symmetry, whenever we obtain a single $S$, all SLMs of the form $R_BSR_A^{-1}$ are known, where $R_\alpha$ is an element of $G_\alpha^\text{rot}$, the group consisting of all rotations that have appeared in the space group of $\alpha$~\cite{SM}.
This accelerates the enumeration by up to $|G_A^\text{rot}|\times|G_B^\text{rot}|$ times (e.g., 576-fold faster when exploring the martensitic transformation of steel which have $|G_A^\text{rot}|=|G_B^\text{rot}|=24$) and simplifies the results.
The enumerated CSMs can be directly used to study concerted MEPs.
With certain geometric criteria~\cite{capillas2007maximal,chen2016determination,stevanovic2018predicting,therrien2020matching}, one can screen out better candidates from enumerated CSMs.
This systematically improves the CSM specification in methods like SSNEB.
On the other hand, metadynamics simulations have found several nucleation MEPs, which proceed locally through CSMs with small $\tilde Z$'s~\cite{laio2002escaping,martovnak2003predicting,badin2021nucleating}.
Our scheme can provide a comprehensive list of CSMs covering this range, not only reproducing nucleation-favored CSMs, but also revealing their distinctiveness among all candidates.
Additionally, a high-throughput computation~\cite{jain2013commentary,saal2013materials,curtarolo2012aflow}---the paradigm on the tide---for SSPTs is also enabled, as we presented a framework to describe, classify and enumerate all CSMs.
The numerous candidates might contain currently unknown yet realistic mechanisms, which can shed light on the design of more predictive screening criteria, and inspire novel understandings of SSPTs both in nature and in industries.
\begin{acknowledgments}
We acknowledge helpful discussions with J.~X.~Zeng, H.~Y.~Yang, R.~C.~He, J.~C.~Gao and W.~J.~Han.
We are grateful to F.~Therrien and Z.~P.~Liu for helpful comments.
We are supported by the National Science Foundation of China under Grants No.~12234001, No.~11934003, No.~12204015, and No.~62321004, the National Basic Research Program of China under Grants No.~2021YFA1400503 and No.~2022YFA1403500, the Beijing Natural Science Foundation under Grant No.~Z200004, and the Strategic Priority Research Program of the Chinese Academy of Sciences Grant No.~XDB33010400.
The computational resources were provided by the supercomputer center at Peking University, China.
\end{acknowledgments}

\end{document}


\title{Supplemental Material:\\Crystal-Structure Matches in Solid-Solid Phase Transitions}
\author{Fang-Cheng Wang}
\affiliation{State Key Laboratory for Artificial Microstructure and Mesoscopic Physics, Frontier Science Center for Nano-optoelectronics, School of Physics, Peking University, Beijing 100871, People's Republic of China}
\author{Qi-Jun Ye}
\email{qjye@pku.edu.cn}
\affiliation{State Key Laboratory for Artificial Microstructure and Mesoscopic Physics, Frontier Science Center for Nano-optoelectronics, School of Physics, Peking University, Beijing 100871, People's Republic of China}
\affiliation{Interdisciplinary Institute of Light-Element Quantum Materials, Research Center for Light-Element Advanced Materials, and Collaborative Innovation Center of Quantum Matter, Peking University, Beijing 100871, People's Republic of China}
\author{Yu-Cheng Zhu}
\affiliation{State Key Laboratory for Artificial Microstructure and Mesoscopic Physics, Frontier Science Center for Nano-optoelectronics, School of Physics, Peking University, Beijing 100871, People's Republic of China}
\author{Xin-Zheng Li}
\email{xzli@pku.edu.cn}
\affiliation{State Key Laboratory for Artificial Microstructure and Mesoscopic Physics, Frontier Science Center for Nano-optoelectronics, School of Physics, Peking University, Beijing 100871, People's Republic of China}
\affiliation{Interdisciplinary Institute of Light-Element Quantum Materials, Research Center for Light-Element Advanced Materials, and Collaborative Innovation Center of Quantum Matter, Peking University, Beijing 100871, People's Republic of China}
\affiliation{Peking University Yangtze Delta Institute of Optoelectronics, Nantong, Jiangsu 226010, People's Republic of China}
\begin{abstract}
This Supplemental Material consists of three parts.
In Section~\ref{results}, we elaborate on the elements directly related to the main text, including how an orientation relationship (OR) is determined by a CSM and the details of enumeration.
In Section~\ref{formalism}, we mathematically rigorize the concepts introduced in the main text, e.g., the CSM, SLM, and periodic $\tilde Z$-atom correspondence.
The uniqueness and finiteness of SLM are also proved.
An additional discussion of geometric criteria is provided in Section~\ref{geo}, which explains the adoption of RMSS and RMSD.
\end{abstract}
\maketitle
\begin{figure*}[!ht]
  \centering
  \includegraphics[width=\linewidth]{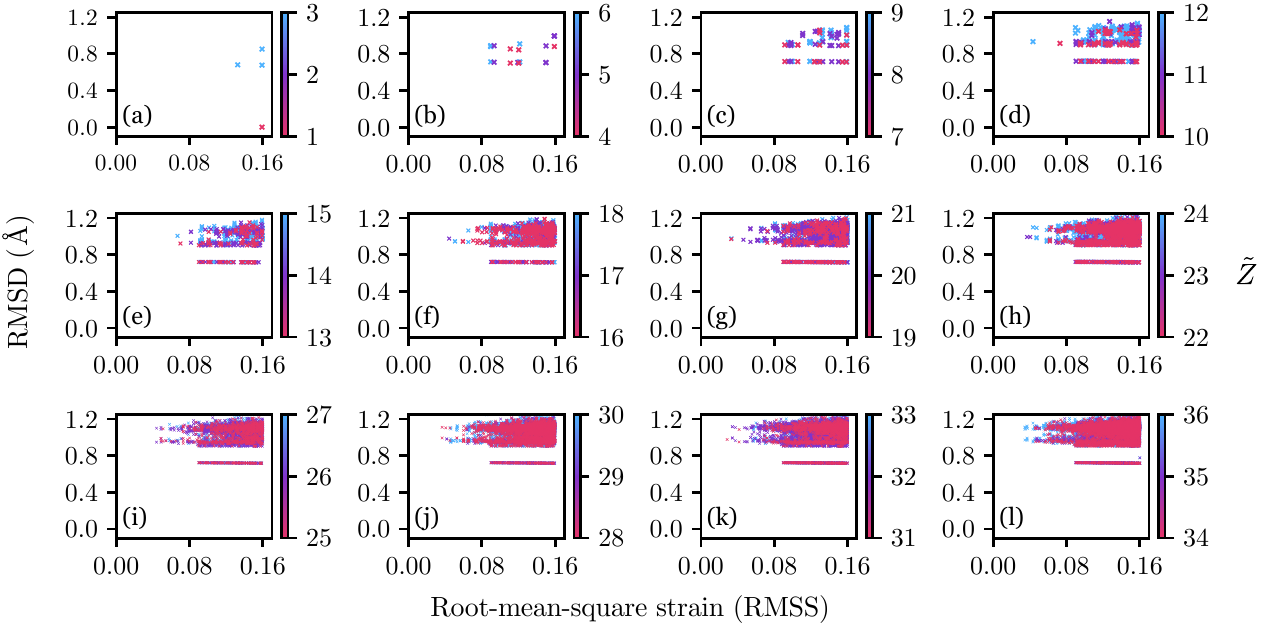}
  \caption{CSMs of the martensitic transformation of steel. For simplicity, only the representative CSM of each SLM is shown.}
  \label{FigS1}
\end{figure*}
\begin{table*}[!ht]
\caption{\label{tab:tables1}%
	Total number of SLMs with $\mathrm{RMSS}\le 16\%$ in the martensitic transformation of steel. Each SLM contributes only one count to the smallest $\tilde Z$ compatible with it, and equivalent SLMs (see Eq.~(\ref{equiv})) are counted only once.
}
\begin{ruledtabular}
\begin{tabular}{ccccccccccccccccccc}
	$\tilde Z$ & 1 & 2 & 3 & 4 & 5 & 6 & 7 & 8 & 9 & 10 & 11 & 12 & 13 & 14 & 15 & 16 & 17 & 18\\
	No.~of SLMs & 1 & 7 & 16 & 40 & 40 & 119 & 73 & 215 & 206 & 336 & 186 & 849 & 267 & 686 & 868 & 1361 & 504 & 2251\\
	\hline&\\[-2.5ex]
	$\tilde Z$ & 19 & 20 & 21 & 22 & 23 & 24 & 25 & 26 & 27 & 28 & 29 & 30 & 31 & 32 & 33 & 34 & 35 & 36\\
	No.~of SLMs & 651 & 2914 & 1886 & 2169 & 1090 & 7232 & 1980 & 3405 & 3641 & 7248 & 2000 & 13313 & 2398 & 13074 & 6225 & 7645 & 6487 & 28412\\
\end{tabular}
\end{ruledtabular}
\end{table*}
\tableofcontents
\section{Enumeration results}\label{results}
\subsection{Martensitic transformation in steel}\label{or}
\begin{figure}[!ht]
	\centering
	\includegraphics[width=\linewidth]{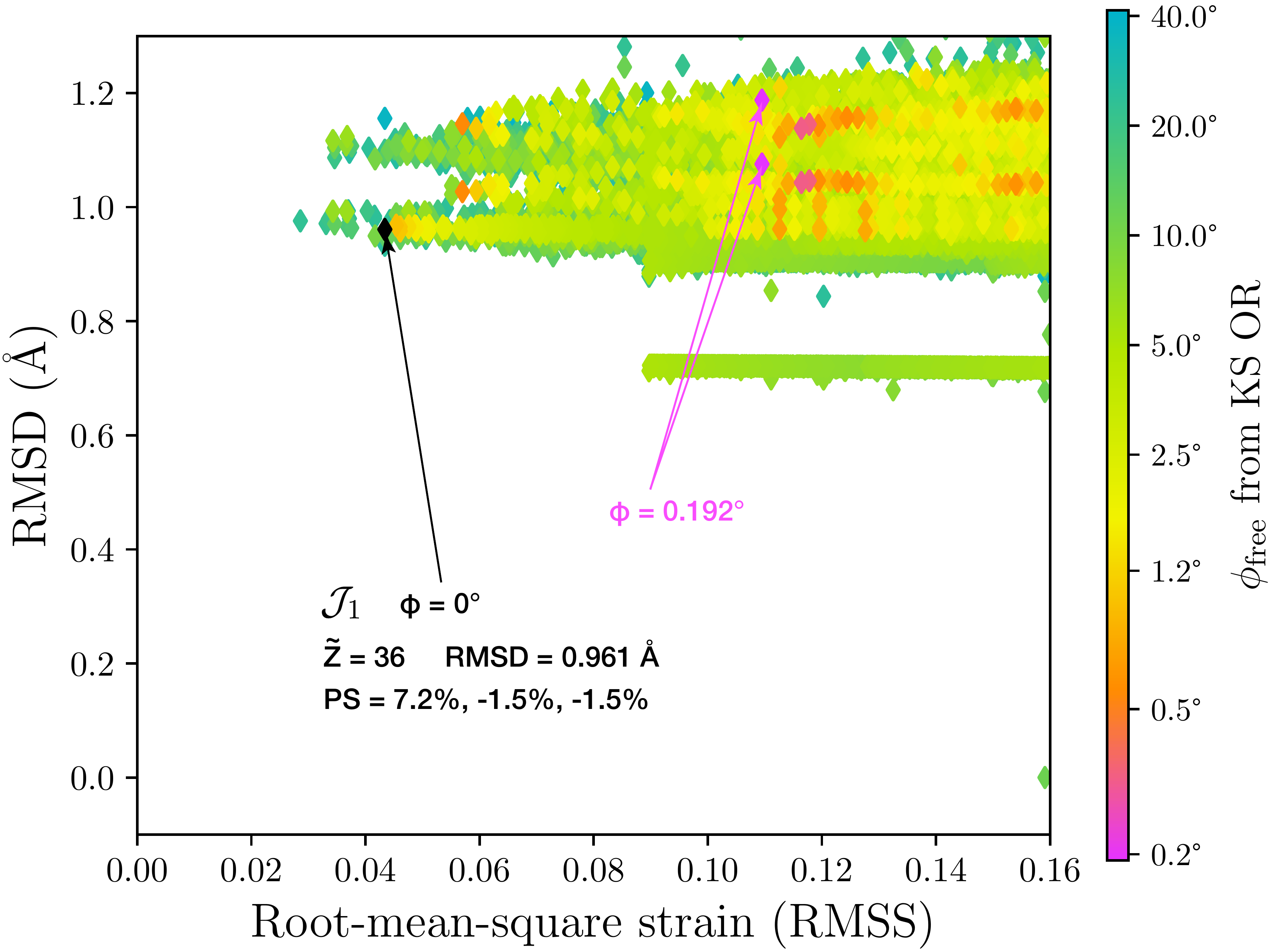}
	\caption{CSMs benchmarked by $\phi_\text{free}$ from the KS OR.}
  \label{FigS2}
\end{figure}

\begin{figure}[!ht]
	\centering
	\includegraphics[width=\linewidth]{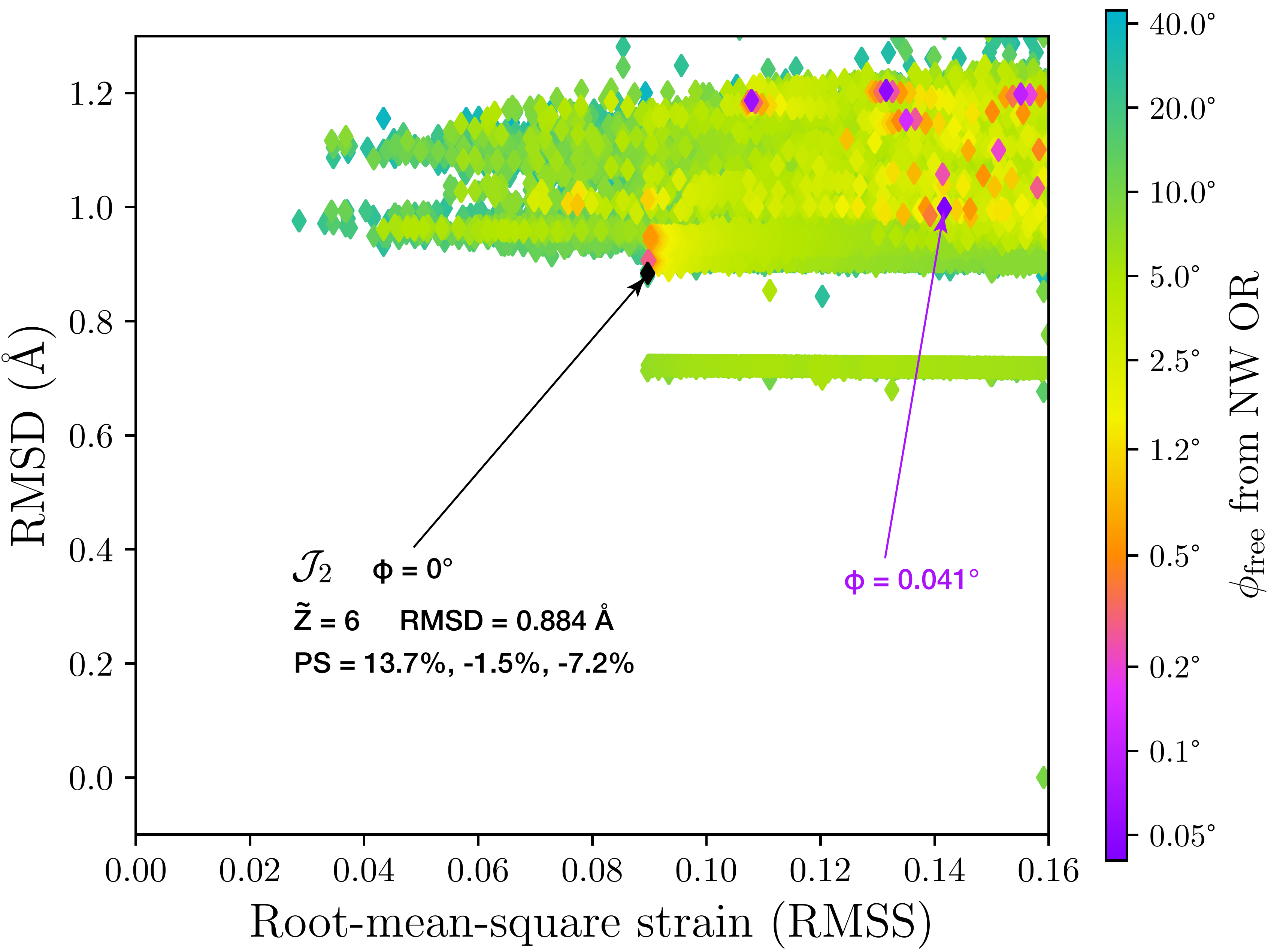}
	\caption{CSMs benchmarked by $\phi_\text{free}$ from the NW OR.}
  \label{FigS3}
\end{figure}

\begin{figure}[!ht]
	\centering
	\includegraphics[width=\linewidth]{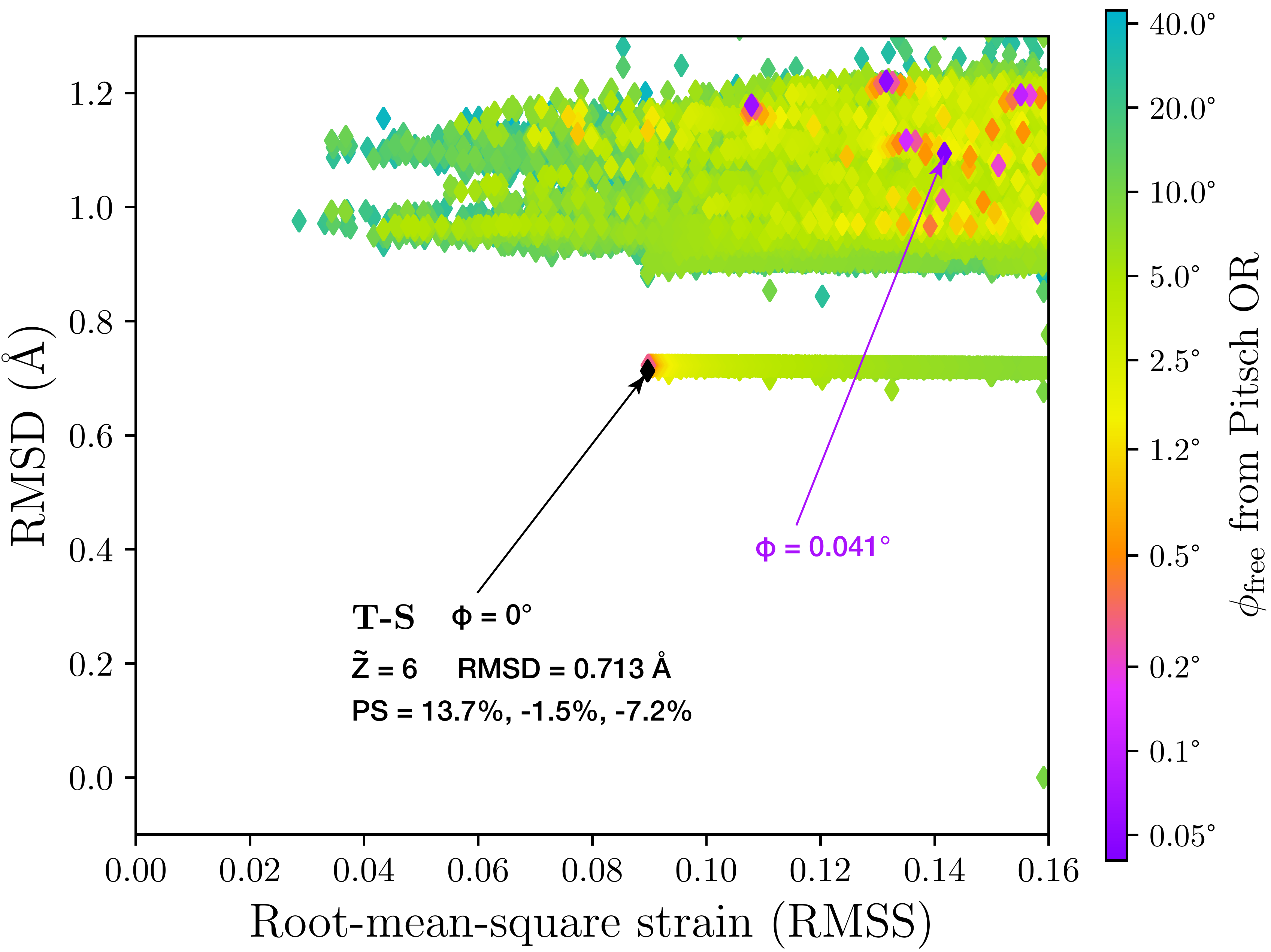}
	\caption{CSMs benchmarked by $\phi_\text{free}$ from the Pitsch OR.}
  \label{FigS4}
\end{figure}

The martensitic transformation of steel is a monatomic SSPT from the austenite phase (fcc, $a_\text{fcc}=3.57\,\text\AA$) to the martensite phase (bcc, $a_\text{bcc}=2.87\,\text\AA$).
To the best of our knowledge, there are only two previously proposed CSMs: the Bain one~\cite{bain1924nature} and the Therrien-Stevanović (T-S) one~\cite{therrien2020minimization}.
Although some other shear mechanisms have been proposed, they all have the same CSM as the Bain mechanism~\cite{therrien2020minimization}.
We hope to find CSMs with root-mean-square strain (RMSS) or displacement (RMSD) smaller than known mechanisms.
Given that the Bain mechanism has $\mathrm{RMSS}=15.91\%$ and $\mathrm{RMSD}=0$, any CSM with $\mathrm{RMSS}>16\%$ is not in consideration since both its RMSS and RMSD are bigger than the Bain mechanism.
On the other hand, CSMs reported in analogous nucleation generally have $\tilde Z\le 8$~\cite{khaliullin2011nucleation,badin2021nucleating,santos2022size}.
Therefore, we exhaust SLMs using our Python package {\sc crystmatch} with safe protocols as $\tilde Z\le 36$ and $\mathrm{RMSS}\le 16\%$.
The representative CSM of each SLM is obtained via the Hungarian algorithm~\cite{kuhn1955hungarian}.
These results are detailed in Fig.~\ref{FigS1} and Table.~\ref{tab:tables1}.
In the rest part of this subsection, we elaborate on how an orientation relationship (OR) is determined by a CSM.
Given the initial structure $\mathcal{A}$, the final structure produced by SSPT may not be strictly $\mathcal{B}$, but can also differ by a rotation and a translation, as
\begin{equation}\label{affine}
	R\mathcal{B}+\boldsymbol{\tau}=\{R\mathbf{r}+\boldsymbol{\tau}\mid \mathbf{r}\in\mathcal{B}\}
,\end{equation}
where $R\in\operatorname{SO}(3)$ and $\boldsymbol{\tau}\in\mathbb{R}^3$ are arbitrary.
An OR is a relative orientation between the initial and final structure described by $R$.
However, for some $R_B\in\operatorname{SO}(3)$, final structures with orientation $R$ and $RR_B$ are indistinguishable in experiments.
These $R_B$'s from a finite subgroup of $\operatorname{SO}(3)$, namely
\begin{equation}
	G_B^\text{rot}=\left\{R\,\big|\,[R|\mathbf{t}]\in G_B\right\}
,\end{equation}
where $G_B$ is the space group of $\mathcal{B}$.
In the same sense, neither does left-multiplying $R_A\in G_A^\text{rot}$ to $R$ produce new \textit{relative} orientation.
Therefore, an experimentally observed OR is not associated with a specific $R$, but an equivalence class
\begin{equation}
	G_A^\text{rot}RG_B^\text{rot}=\{R_ARR_B\mid R_A\in G_A^\text{rot},R_B\in G_B^\text{rot}\}
.\end{equation}
The conventional notation of OR consists of two or more parallelisms (see Table~\ref{tab:tables2}) which determine an $R$.
Such $R$ varies in different literatures~\cite{koumatos2017theoretical}, since it can be any element of $G_A^\text{rot}RG_B^\text{rot}$.
There are two manners for a CSM with SLM $S$ to \textit{a priori} determine an OR.
As proved in Subsection~\ref{criteria1}, the total distance traveled by the atoms is minimized when $\mathcal{A}$ is deformed by $\sqrt{S^\text{T}S}$, which is rotation-free.
Since $S\mathcal{A}$ is associated with $\mathcal{B}$ via periodic atomic displacement, $\sqrt{S^\text{T}S}\,\mathcal{A}$ is associated with $R_S\mathcal{B}$, where
\begin{equation}\label{fs}
  R_S=\sqrt{S^\text{T}S}\,S^{-1} 
,\end{equation}
is the rotation-free orientation determined by $S$.
For each SLM $S$, its distinction from the experimentally observed OR $R$ is quantified by the least rotation angle required to produce an element in $G_A^\text{rot}RG_B^\text{rot}$, namely
\begin{equation}\label{thfree}
	\phi_\text{free}(S)=\min_{R'\in G_A^\text{rot}RG_B^\text{rot}}\theta\!\left[(R_S)^{-1}R'\right]
,\end{equation}
where
\begin{equation}
	\theta(X)=\arccos\frac{\operatorname{Tr}(X)-1}{2}
\end{equation}
is the rotation angle of $X\in\operatorname{SO}(3)$.
We score the enumerated CSMs by Eq.~(\ref{thfree}), as shown in Figs.~\ref{FigS2}, \ref{FigS3} and \ref{FigS4}.
The phenomenological theory of martensitic transformation (PTMT)~\cite{khachaturyan2013theory} determines the OR in another manner, which assumes that the habit plane---the experimentally observed phase interface---is an \textit{invariant plane} under the lattice deformation.
Such plane consists of the vectors whose lengths are invariant, and can be formally defined as
\begin{equation}\label{iplane}
	  \mathcal{U}^*=\big\{\mathbf{v}\,\big|\,\mathbf{v}\in\mathbb{R}^3,|S\mathbf{v}|=|\mathbf{v}|\big\}
.\end{equation}
Denoting the $j$-th largest singular value~\cite{axler1997linear} of $S$ by $s_j$, Eq.~(\ref{iplane}) is two-dimensional only if $s_2=1$.
This is only possible when a principal strain (PS) of $S$ is \textit{zero}, which never strictly happens when experimental lattice constants are used.
\begin{figure}[!ht]
  \centering
  \includegraphics[width=\linewidth]{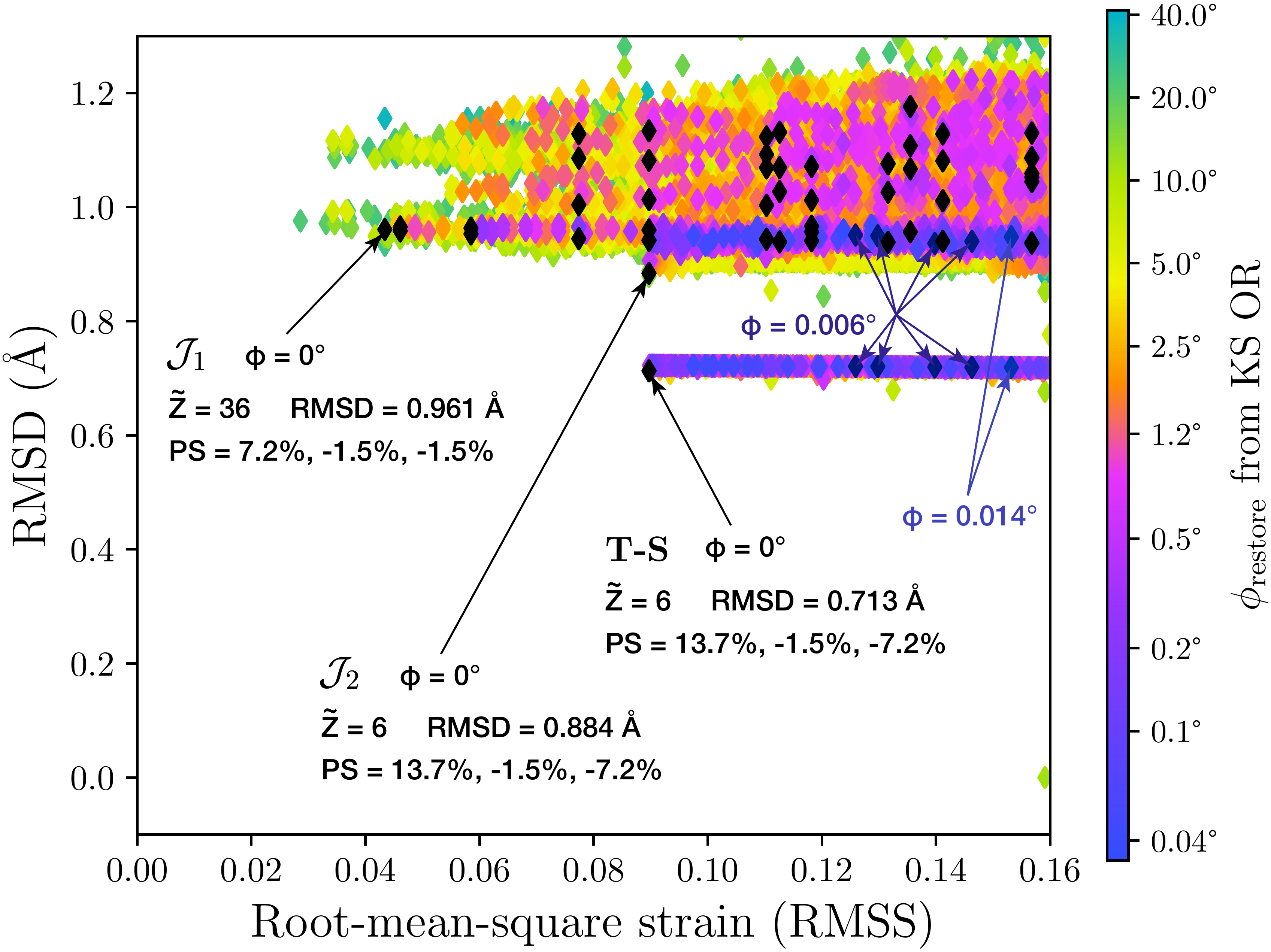}
	\caption{CSMs benchmarked by $\phi_\text{restore}$ from the KS OR.}
  \label{FigS5}
\end{figure}
\begin{figure}[!ht]
  \centering
  \includegraphics[width=\linewidth]{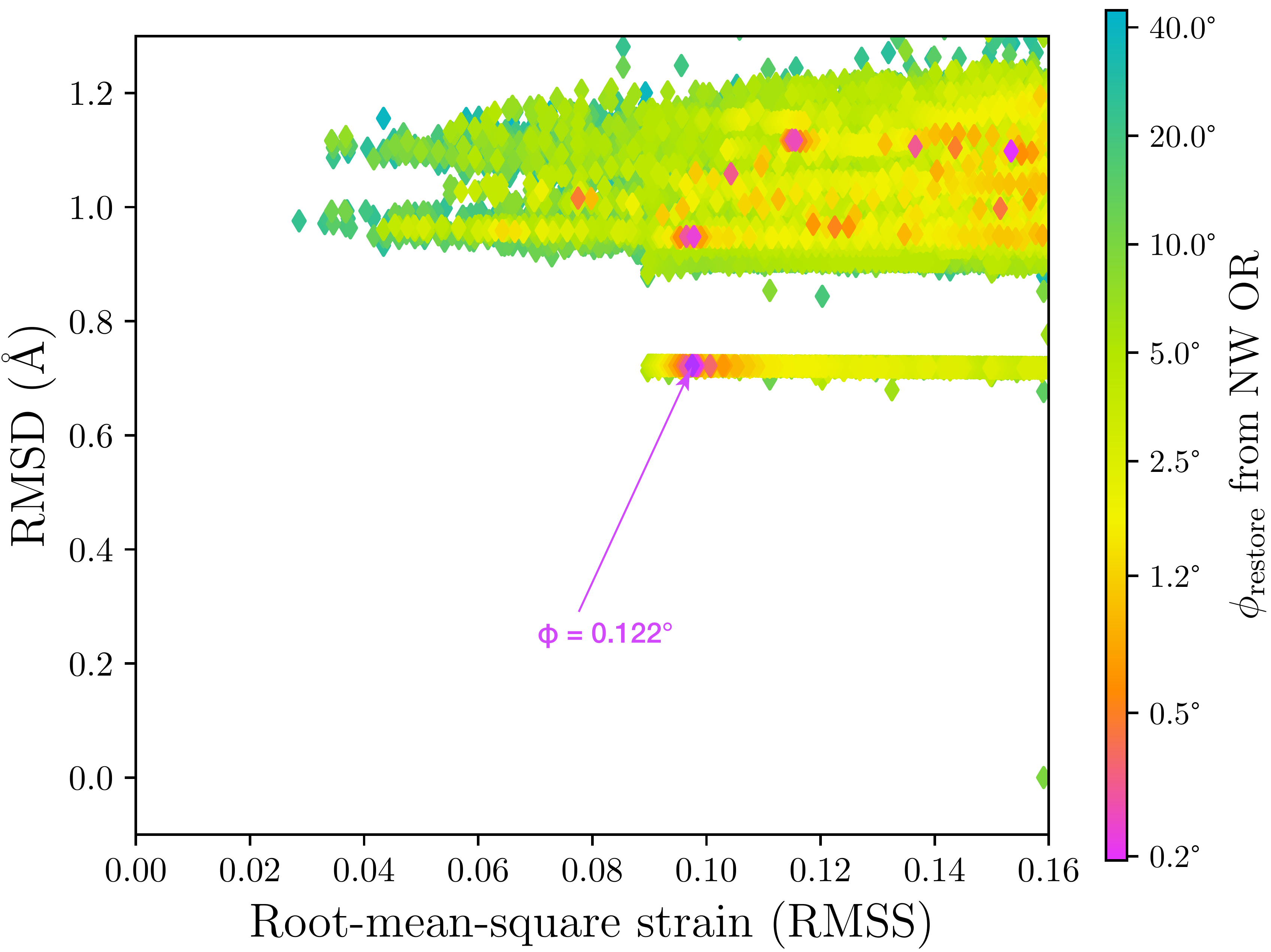}
	\caption{CSMs benchmarked by $\phi_\text{restore}$ from the NW OR.}
  \label{FigS6}
\end{figure}
\begin{figure}[!ht]
  \centering
  \includegraphics[width=\linewidth]{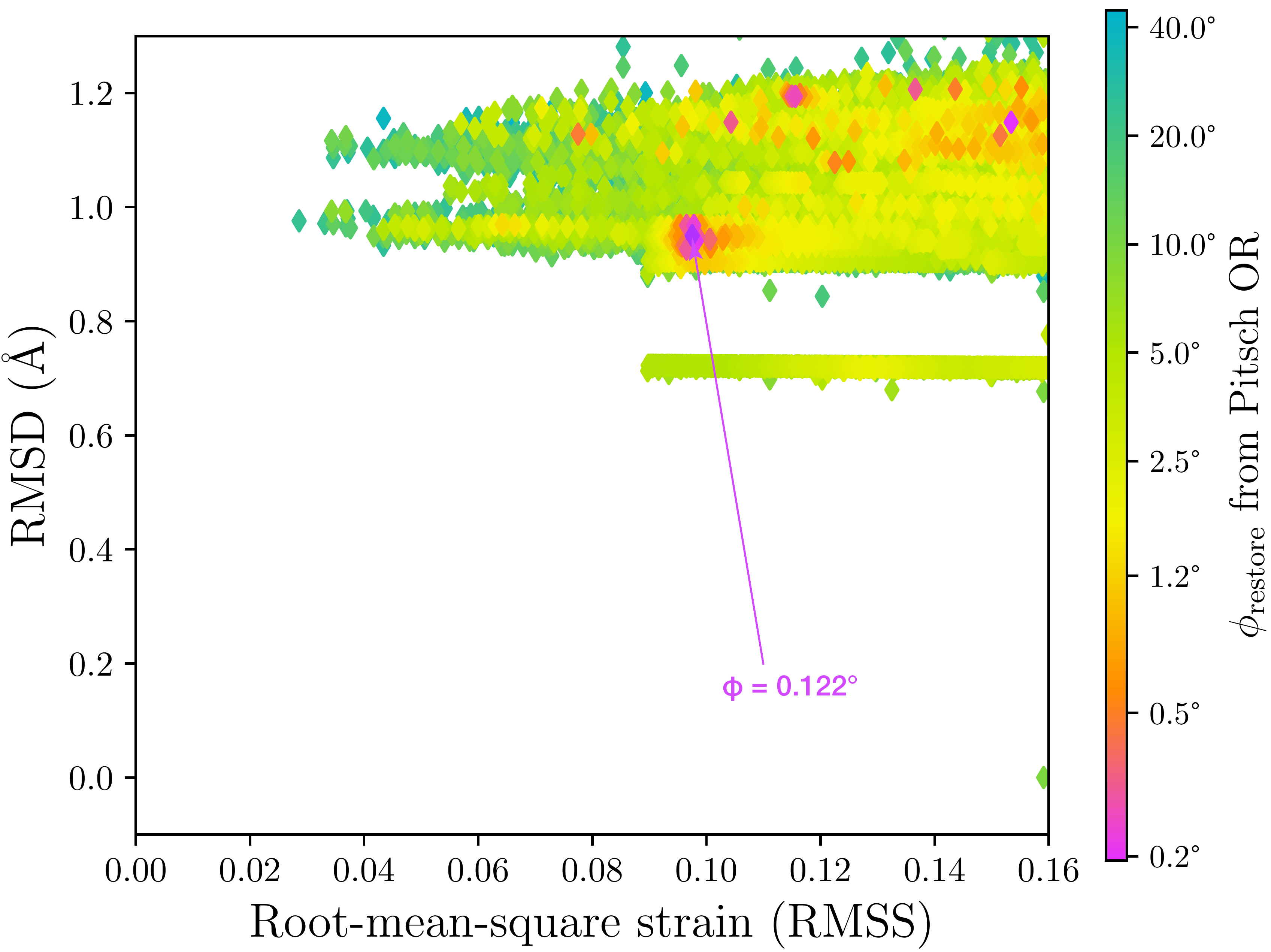}
	\caption{CSMs benchmarked by $\phi_\text{restore}$ from the Pitsch OR.}
  \label{FigS7}
\end{figure}

\begin{table}[b]
\caption{\label{tab:tables2}%
	Experimentally observed ORs in the martensitic transformation of steel, according to Ref.~\cite{koumatos2017theoretical}.
}
\begin{ruledtabular}
\begin{tabular}{lcc}
OR&
Parallelism 1&
Parallelism 2\\
\hline &\\[-2.5ex]
KS& $(111)_\text{fcc}\parallel(011)_\text{bcc}$ & $[01\bar 1]_\text{fcc}\parallel[1\bar 11]_\text{bcc}$\\
NW& $(111)_\text{fcc}\parallel(011)_\text{bcc}$ & $[01\bar 1]_\text{fcc}\parallel[\bar 100]_\text{bcc}$\\
Pitsch& $(110)_\text{fcc}\parallel(\bar 1\bar 1\bar 2)_\text{bcc}$ & $[001]_\text{fcc}\parallel[1\bar 10]_\text{bcc}$\\
\end{tabular}
\end{ruledtabular}
\end{table}
\begin{figure*}[!ht]
	\centering
	\includegraphics[width=0.6\linewidth]{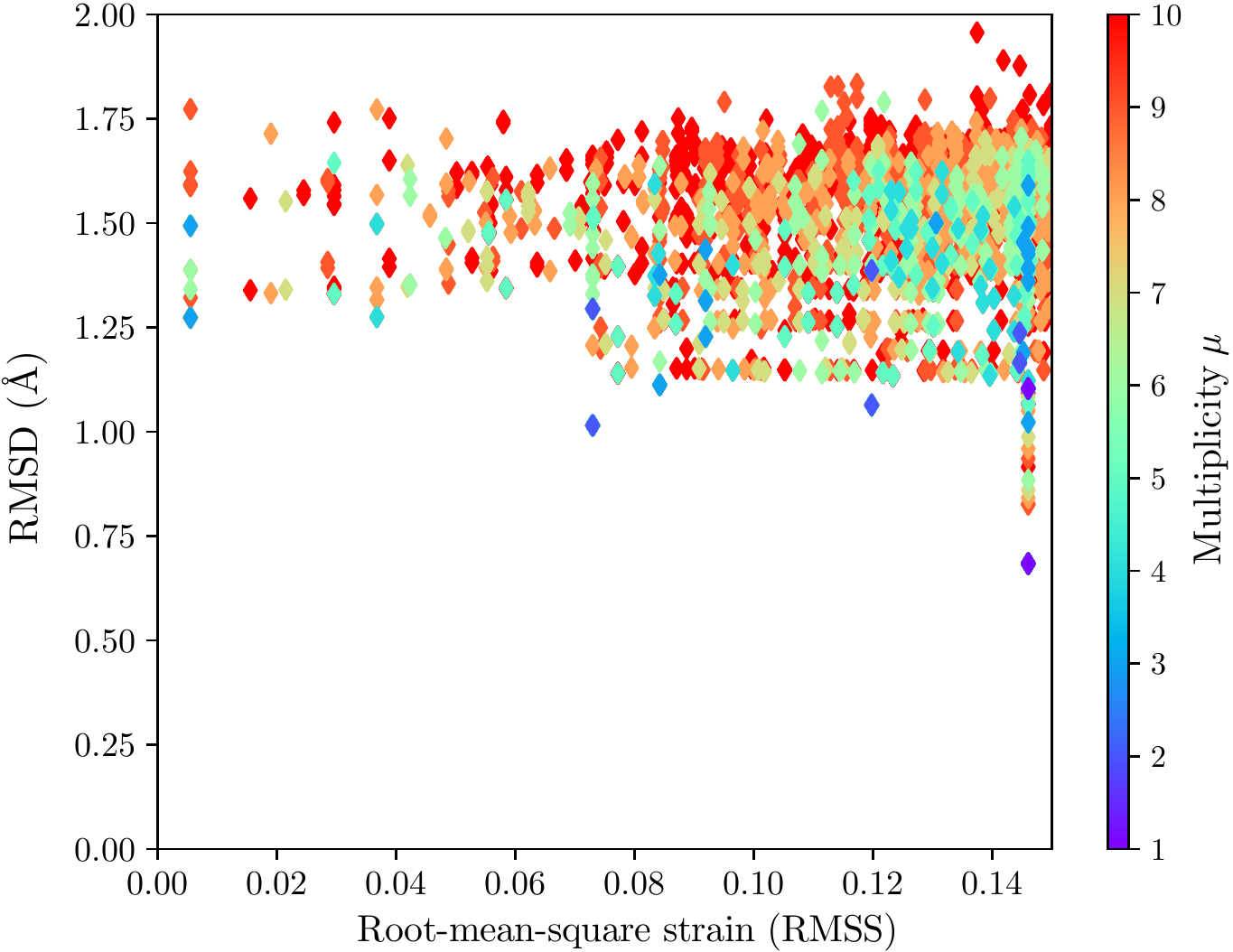}
	\caption{CSMs of the B4-B3 transition in ZnS, enumerated in the range $\mathrm{RMSS}\le 15\%$, $\mu\le 10$. For simplicity, only the representative CSM of each SLM is shown. The color of the marks indicates the CSM multiplicity $\mu$, while we have $\tilde Z=4\mu$ in this case.}
  \label{FigS8}
\end{figure*}
To address this issue, the PTMT turns to assume that the habit plane is a \textit{uniformly scaled plane} (USP)
\begin{equation}
  \mathcal{U}=\big\{\mathbf{v}\,\big|\,\mathbf{v}\in\mathbb{R}^3,|S\mathbf{v}|=k|\mathbf{v}|\big\}
,\end{equation}
where $k$ is some positive number.
When $s_1=s_2=s_3$, we have $\mathcal{U}=\mathbb{R}^3$.
In other cases, denoting the orthonormal eigenvectors of $\sqrt{S^\text{T}S} $ (right singular vectors of $S$) by $\mathbf{e}_1$, $\mathbf{e}_2$, and $\mathbf{e}_3$, we have $k=s_2$ and $\mathcal{U}=\mathcal{U}_+\cup\mathcal{U}_-$, where~\cite{therrien2020minimization}
\begin{equation}\label{uplane}
	\mathcal{U}_\pm=\operatorname{span}\!\left\{\mathbf{e}_2,\sqrt{ s_2^2- s_3^2}\,\mathbf{e}_1\pm \sqrt{ s_1^2- s_2^2}\,\mathbf{e}_3 \right\}
\end{equation}
is indeed two-dimensional.
To derive the USP-restoring orientation relationship, we consider its difference from the rotation-free one.
Under the rotation-free deformation $\sqrt{S^\text{T}S} $, the USP $\mathcal{U}_\pm$ evolves to
\begin{equation}\label{uplaned}
	\sqrt{S^\text{T}S}\,\mathcal{U}_\pm=\operatorname{span}\!\left\{\mathbf{e}_2,s_1 \sqrt{s_2^2-s_3^2}\,\mathbf{e}_1\pm s_3 \sqrt{s_1^2-s_2^2}\,\mathbf{e}_3\right\}
,\end{equation}
as indicated by the pink planes in Fig.~4 of the main text.
We note that Eq.~(\ref{uplaned}) also equals $R_\pm \mathcal{U}_\pm$, where $R_\pm$ is a rotation with $\mathbf{e}_2$ as the axis by an angle of
\begin{equation}
  \phi_\pm=\pm\arctan \frac{\sqrt{s_1^2-s_2^2} \sqrt{s_2^2-s_3^2}}{s_1s_3+s_2^2}
.\end{equation}
In other words, to restore $\mathcal{U}_\pm$, the final structure $R_S\mathcal{B}$ must undergo an additional rotation $R_\pm^{-1}$.
Therefore, the distinction between SLM $S$ and OR $R$ is quantified by
\begin{equation}
	\phi_\text{restore}(S)=\min_{\substack{R'\in G_A^\text{rot} R G_B^\text{rot}\\R''\in\{R_+,R_-\}}}\theta\!\left[R''(R_S)^{-1}R'\right]
.\end{equation}
So far, we see that in either manner, the OR is alone determined by the SLM of a CSM.
When using the USP-restoring manner, 55 CSMs can reproduce the KS OR (black diamonds in Fig.~\ref{FigS5}), while \textit{none} of the enumerated CSMs can produce the NW or Pitsch OR (see Figs.~\ref{FigS6} and \ref{FigS7}).
We therefore believe that the habit plane undergoes surface reconstruction to cope with the lattice mismatch when $s_2\neq 1$, making the habit plane no longer the USP.
It is also worth noting that $\mathcal{J}_1$ reproduces the KS OR in either manner, i.e., the two USPs of $\mathcal{J}_1$ coincide and stay still without additional rotations $R_\pm^{-1}$ (see Fig.~4(a) in the main text).
This is a consequence of $s_2= s_3$.
Moreover, if close-packing lattice constants ($a_\text{bcc}=\sqrt{2/3}\,a_\text{fcc}$) are used, we have $s_2=s_3=1$.
In this case, the USP of $\mathcal{J}_1$ is an \textit{invariant plane} and do \textit{not} rotate under $\sqrt{S^\text{T}S} $, which is even more demanding than that required by Eq.~(\ref{iplane}).
This may explain the predominance of the KS OR.
\subsection{B4-B3 transition in ZnS}\label{ZnS}
We choose ZnS as a polyatomic system to demonstrate our scheme.
The SSPT from the wurtzite phase (B4, $a=3.81$, $c=6.23$) to the zincblende phase (B3, $a=5.42$) is investigated.
Unlike the martensitic transformation of steel, the B4-B3 transition cannot be done by pure lattice deformation, since no SLM $S$ can make $S\mathcal{A}=R\mathcal{B}+\boldsymbol{\tau}$.
Therefore, we use $\mathrm{RMSS}\le 15\%$ as a moderate bound and perform a fast enumeration with $\mu\le 10$, where $\mu$ is the multiplicity of CSM defined in Eq.~(\ref{mu}).
As shown in Fig.~\ref{FigS8}, no CSM is as significantly superior to the Bain mechanism in terms of RMSD.
However, several CSMs with very small strains are discovered, whose multiplicities are all 3, 6, or 9.
Fig.~\ref{FigS8} shares some common features with Fig.~1 in the main text: i) No CSM stands out in terms of both strain and atomic displacement; ii) The CSM with the lowest RMSD usually has a low multiplicity; iii) Many different SLMs have the exact same strain, especially the black diamonds in Fig.~\ref{FigS5}; iv) Many different CSMs have the exact same RMSD.
These interesting patterns need explanation and further verification.
\section{Mathematical formalism}\label{formalism}
In the main text, we have discussed the CSM, SLM, periodic $\tilde Z$-atom correspondence, supercell pair and IMT representations.
In this section, we formally introduce these concepts, and define the equivalence relation on the set of SLMs.
The finiteness of SLMs within arbitrary strain bound is also proved.
\begin{table*}[!ht]
\caption{\label{tab:tables3}%
	Notations frequently used in Subsection~\ref{csm}.
}
	\begin{tabular}{>{\centering\arraybackslash}p{1.6cm}p{12cm}}
\hline\hline
Notation & Meaning \\
\hline&\\[-2.5ex]
$\mathcal{A}$&Crystal structure, an infinite set consisting of atomic positions\\
$L_A$&Lattice, a full rank translation group of crystal structure $\mathcal{A}$\\
$C_A$&Primitive-cell matrix, whose columns generate $L_A$\\
$Z_{A}$&Number of atoms in $C_A$, equal to the size of the quotient set $|\mathcal{A} / L_A|$\\
$\tilde C_A$&Supercell matrix, whose columns generate a sublattice of $L_A$\\
\hline&\\[-2.5ex]
$\mathcal{J}$&Crystal structure match, a bijection from $\mathcal{A}$ to $\mathcal{B}$\\
$\tilde L$&Lattice of a \textit{periodic} CSM, necessarily a common sublattice of $L_A$ and $L_B$\\
$\tilde C$&Common-supercell matrix, whose columns generate $\tilde L$\\
$\tilde Z$&Number of atoms in $\tilde C$, equal to $|\mathcal{A} / \tilde L|=|\mathcal{B} / \tilde L|$, a common multiple of $Z_A$ and $Z_B$\\
\hline&\\[-2.5ex]
$S$&Sublattice match of $\mathcal{J}$, which makes $\mathcal{J}^S$ a periodic CSM\\
$S\mathcal{A}$&Crystal structure deformed by $S$, having a common sublattice with $\mathcal{B}$\\
$\mathcal{J}^S$&Periodic CSM between $S\mathcal{A}$ and $\mathcal{B}$, deformed from a non-periodic $\mathcal{J}$ via $S$\\
\hline\hline
\end{tabular}
\end{table*}
\subsection{Crystal-structure match}\label{csm}
A CSM $\mathcal{J}$ is a one-to-one map between two infinite sets of atoms $\mathcal{A}$ and $\mathcal{B}$.
However, such infinite map is not convenient to describe, simulate, analyze and optimize.
Using the translational symmetry, it is possible to describe $\mathcal{J}$ by a finite number of 3D vectors.
This can be easily done when $\mathcal{J}$ is \textit{periodic}, i.e., with a full rank translation group $\tilde L$.
The role of SLM is to deform a non-periodic CSM into a periodic one.
We shall rigorize these concepts in the following, where the system is assumed to be monatomic for simplicity.
The generalization to polyatomic SSPTs is self-evident, whose validity has been shown in Subsection~\ref{ZnS}.
Notations frequently used in this subection are summarized in Table.~\ref{tab:tables3}.
We formally define a structure $\mathcal{A}$ as a countable subset of $\mathbb{R}^3$ without accumulation points.
A translation element of $\mathcal{A}$ is a vector $\mathbf{t}\in \mathbb{R}^3$ such that after its translation, $\mathcal{A}+\mathbf{t}=\left\{\mathbf{a}+\mathbf{t}\mid \mathbf{a}\in\mathcal{A}\right\}$ is the same as $\mathcal{A}$.
We consider only \textit{crystal} structures, whose translation elements form a lattice, i.e., a full rank additive group of vectors, denoted by
\begin{equation}
  L_A=\left\{\mathbf{t}\,\big|\, \mathcal{A}+\mathbf{t}=\mathcal{A}\right\}
.\end{equation}
A \textit{basis} of $L_A$ consists of three vectors generating $L_A$ via integer linear combination, which can be organized as column vectors (in right-handed order) to form a $3\times 3$ matrix $C_A=[\mathbf{t}_1,\mathbf{t}_2,\mathbf{t}_3]$.
Each $C_A$ is associated with a primitive cell in $\mathcal{A}$, whose volume is $\det C_A$.
We say two atoms in $\mathcal{A}$ are translational equivalent, if they are one translation element away from each other.
By the definition of crystal structure, the quotient set $\mathcal{A}/L_A$ induced by such equivalence relation is finite (a consequence of the Bolzano-Weierstrass theorem~\cite{bartle2000introduction}).
We denote the set size $|\mathcal{A} /L_A|$ by $Z_A$, which equals the number of (inequivalent) atoms in a primitive cell.
For crystal structures $\mathcal{A}$ and $\mathcal{B}$ of identical chemical composition, we define a CSM between $\mathcal{A}$ and $\mathcal{B}$ as a bijection $\mathcal{J}\colon\mathcal{A}\to\mathcal{B}$.
From the set-theoretic point of view, $ \mathcal{J}$ can also be viewed as a set of \textit{atom pairs}, i.e., a subset of $\mathcal{A}\times\mathcal{B}$ such that
\begin{align}
	\forall \mathbf{a}\in\mathcal{A},\quad \exists !(\mathbf{x},\mathbf{y})\in \mathcal{J},\quad \mathbf{x}&=\mathbf{a},\label{atob}\\
	\forall \mathbf{b}\in\mathcal{B},\quad \exists !(\mathbf{x},\mathbf{y})\in \mathcal{J},\quad \mathbf{y}&=\mathbf{b}\label{btoa}
.\end{align}
In other words, each atom in $\mathcal{A}$ appears exactly once in $ \mathcal{J}$, and so do those atoms in $\mathcal{B}$.
Given a CSM $\mathcal{J}$ between $\mathcal{A}$ and $\mathcal{B}$, if there exists a full rank additive group of vectors
\begin{equation}
  \tilde L=\left\{\mathbf{t}\,\big|\,\mathcal{J}+\mathbf{t}=\mathcal{J}\right\}
,\end{equation}
where
\begin{equation}\label{periodj}
	\mathcal{J}+\mathbf{t}=\left\{(\mathbf{x}+\mathbf{t},\mathbf{y}+\mathbf{t})\,\big|\, (\mathbf{x},\mathbf{y})\in \mathcal{J}\right\}
,\end{equation}
we say that $ \mathcal{J}$ is \textit{periodic} and $\tilde L$ is the lattice of $\mathcal{J}$.
From Eq. (\ref{periodj}) we see that each $\mathbf{t}\in\tilde L$ is a translation element of $\mathcal{A}$, and the same is true for $\mathcal{B}$.
Therefore, $\tilde L$ is a common sublattice of $L_A$ and $L_B$.
As long as $ \mathcal{J}$ is periodic, there is a canonical bijection $\tilde{ \mathcal{J}}$ between $\mathcal{A} /\tilde L$ and $\mathcal{B} /\tilde L$, namely
\begin{align}
	\forall \mathbf{a}\in\mathcal{A},\quad \tilde{ \mathcal{J}}(\mathbf{a}+\tilde L)&= \mathcal{J}(\mathbf{a})+\tilde L,\\
	\forall\mathbf{b}\in\mathcal{B},\quad\tilde{ \mathcal{J}}^{-1}(\mathbf{b}+\tilde L)&= \mathcal{J}^{-1}(\mathbf{b})+\tilde L
,\end{align}
where $\mathbf{a}+\tilde L=\{\mathbf{a}+\mathbf{t}\mid \mathbf{t}\in\tilde L\}$ is the element in $\mathcal{A} /\tilde L$ to which $\mathbf{a}\in\mathcal{A}$ belongs.
The basis of $\tilde L$ forms a $3\times 3$ matrix $\tilde C$ associated with a supercell containing finitely many atoms.
Therefore, both $\mathcal{A} /\tilde L$ and $\mathcal{B} /\tilde L$ are finite, and the existence of a bijection between them means that they are the same size, which we denote be $\tilde Z$.
So far, we see that if $ \mathcal{J}$ is periodic, then $\mathcal{A}$ and $\mathcal{B}$ must have the same number density $\tilde Z /\det \tilde C$.
In SSPTs, the initial and final crystal structures generally have different densities, in which case $ \mathcal{J}\colon\mathcal{A}\to\mathcal{B}$ can never be periodic.
However, there may exist a linear transformation $S$ on $\mathbb{R}^3$ such that after its deformation, the induced CSM between $S\mathcal{A}=\left\{S\mathbf{a}\mid \mathbf{a}\in\mathcal{A}\right\}$ and $\mathcal{B}$ is periodic.
Specifically, we define
\begin{equation}
	\mathcal{J}^S=\left\{(S\mathbf{a},\mathbf{b})\,\big|\,(\mathbf{a},\mathbf{b})\in\mathcal{J}\right\}
,\end{equation}
which is a bijection between $S\mathcal{A}$ and $\mathcal{B}$.
When $\mathcal{J}^S$ is a periodic CSM between $S\mathcal{A}$ and $\mathcal{B}$, we call $(S,\tilde L)$ an SLM of $\mathcal{J}$, where $\tilde L$ is the lattice of $\mathcal{J}^S$.
Since $S\mathcal{A}$ must have the same density as $\mathcal{B}$, we have
\begin{equation}\label{dets}
	\det S=\rho_A / \rho_B
,\end{equation}
where $\rho_\alpha = Z_\alpha / \det C_\alpha$.
A CSM has either a \textit{unique} SLM, or no SLM.
Otherwise, consider that both $S_1$ and $S_2$ are SLMs of the same $ \mathcal{J}$.
Denote the lattice of $ \mathcal{J}^{S_1}$ and $ \mathcal{J}^{S_2}$ by $\tilde L_1$ and $\tilde L_2$, respectively.
For an arbitrary $(\mathbf{a},\mathbf{b})\in  \mathcal{J}$, we have
\begin{align}
	\forall \mathbf{t}\in\tilde L_1,\quad
	(S_1\mathbf{a}+\mathbf{t},\mathbf{b}+\mathbf{t})\in \mathcal{J}^{S_1},\\
	\forall \mathbf{t}\in\tilde L_2,\quad
	(S_2\mathbf{a}+\mathbf{t},\mathbf{b}+\mathbf{t})\in \mathcal{J}^{S_2}
.\end{align}
In other words,
\begin{align}
	\forall \mathbf{t}\in\tilde L_1,\quad
	(\mathbf{a}+S_1^{-1}\mathbf{t},\mathbf{b}+\mathbf{t})\in \mathcal{J},\label{slmt1}\\
	\forall \mathbf{t}\in\tilde L_2,\quad
	(\mathbf{a}+S_2^{-1}\mathbf{t},\mathbf{b}+\mathbf{t})\in \mathcal{J}\label{slmt2}
.\end{align}
We can always choose three linearly independent vectors $\mathbf{t}_1,\mathbf{t}_2,\mathbf{t}_3$ from $\tilde L_1\cap\tilde L_2$ (proved later in this subsection), and Eqs.~(\ref{btoa}), (\ref{slmt1}), and (\ref{slmt2}) imply that
\begin{equation}
	S_1^{-1}[\mathbf{t}_1,\mathbf{t}_2,\mathbf{t}_3]=S_2^{-1}[\mathbf{t}_1,\mathbf{t}_2,\mathbf{t}_3]
.\end{equation}
Since the $3\times 3$ matrix $[\mathbf{t}_1,\mathbf{t}_2,\mathbf{t}_3]$ is invertible, we have $S_1=S_2$.
Consider a CSM $\mathcal{J}$ with SLM $(S,\tilde L)$, whose induced periodic CSM $\mathcal{J}^{S}$ has $\tilde Z$ atoms in its spatial period.
Such $\tilde Z$ is a feature of $\mathcal{J}$ that characterizes its intricacy---larger $\tilde Z$ means weaker periodicity of $\mathcal{J}^S$.
In this sense, we regard CSMs that have no SLM, i.e., cannot be transformed into a periodic one, as having $\tilde Z=\infty$.
Since $\tilde Z$ is a common multiple of $Z_A$ and $Z_B$, it takes values in
\begin{equation}\label{mu}
	\tilde Z=\mu \operatorname{lcm}(Z_A,Z_B)
,\end{equation}
where ``lcm'' means the least common multiple.
We call $\mu$ the multiplicity of $\mathcal{J}$, which takes values in all positive integers, classifying all CSMs.
Since $\mathcal{J}^{S}$ is periodic, we have a full rank $\tilde L$, which is a common sublattice of $SL_A=\left\{S\mathbf{t}\mid \mathbf{t}\in L_A\right\}$ and $L_B$.
A basis of $\tilde L$ can either be generated by columns of $SC_A$, or by columns of $C_B$.
In other words, there exist nonsingular $3\times 3$ integer matrices $M_A$ and $M_B$ such that
\begin{equation}\label{commonsublattice}
 SC_AM_A=C_BM_B
.\end{equation}
Hence, an SLM can always be written as
\begin{equation}\label{scp}
	S=C_B M_B M_A^{-1} C_A^{-1}
,\end{equation}
where $M_A$ and $M_B$ satisfy the constraint of Eq.~(\ref{dets}), namely
\begin{equation}
  \frac{\det M_B}{\det M_A}=\frac{\det C_A}{\det C_B}\frac{\rho_A}{\rho_B}=\frac{Z_A}{Z_B}
.\end{equation}
If we require the columns of Eq.~(\ref{commonsublattice}) to be a basis of $\tilde L$, the supercell $SC_AM_A$ must contain exactly $\tilde Z$ atoms in $S\mathcal{A}$, and so does $C_BM_B$ in $\mathcal{B}$.
In this case, we have
\begin{equation}\label{detm}
	\det M_\alpha=\frac{\tilde Z}{Z_\alpha}\pod{\alpha=A\text{ or }B}
.\end{equation}
Given the SLM $(S,\tilde L)$, the only unspecified part of $\mathcal{J}$ is how the $\tilde Z$ atoms in $\tilde C_A$ are mapped to atoms in $\mathcal{B}$.
This uncertainty is twofold: i) $\mathcal{J}(\mathbf{a})$ can belong to any of the $\tilde Z$ translational equivalence classes in $\mathcal{B}/\tilde L$, and ii) there are still infinitely many possible choices in a given equivalence class $\mathbf{b}+\tilde L$.
Fortunately, one may always choose the nearest counterpart for each $S\mathbf{a}$ in aspect ii) to obtain a ``better'' $\mathcal{J}^S$.
As for aspect i), we develop geometric criteria (see Subsection~\ref{criteria2}) to score different $\mathcal{J}^S$, and use the Hungarian algorithm~\cite{kuhn1955hungarian} to optimize it.
Finally, we prove the lemma used in demonstrating the uniqueness of the SLM: given $\tilde L_1$ and $\tilde L_2$ which are both sublattices of $L_B$, their intersection $\tilde L_1\cap\tilde L_2$ is also full rank.
Note that the bases of $\tilde L_1$ and $\tilde L_2$ respectively form supercell matrices $\tilde C_1=C_BM_1$ and $\tilde C_2=C_BM_2$, where $M_1$ and $M_2$ are nonsingular (a consequence of full rank) $3\times 3$ integer matrices.
If there exist nonsingular $3\times 3$ integer matrices $M_3$ and $M_4$ such that
\begin{equation}\label{commont}
  (C_BM_1)M_3=(C_BM_2)M_4
,\end{equation}
then the columns of Eq.~(\ref{commont}) are three linearly independent vectors in $\tilde L_1\cap \tilde L_2$ and the lemma is proved.
Since Eq.~(\ref{commont}) is equivalent to
\begin{equation}
	(M_2^{-1}M_1)M_3=M_4
,\end{equation}
where the elements of $M_2^{-1}M_1$ are rational, we can always let $M_3$ be a scalar matrix whose value is a common multiple of all denominators in $M_2^{-1}M_1$.
Therefore, nonsingular integer matrices $M_3$ and $M_4$ satisfying Eq.~(\ref{commont}) always exist.
\subsection{Unimodular matrix}\label{unimodular}
Given a supercell pair $(M_A,M_B)$ satisfying Eq.~(\ref{detm}), an SLM $(S,\tilde L)$ is completely determined: $S$ is given by Eq.~(\ref{scp}) and $\tilde L$ is generated by the columns of Eq.~(\ref{commonsublattice}).
We call $(M_A,M_B)$ the supercell pair representation of this SLM.
The supercell pair representation of $(S,\tilde L)$ is not unique.
$(M_A,M_B)$ and $(M_A',M_B')$ give the same $S$ if and only if $M_BM_A^{-1}=M_B'M_A'^{-1}$.
On the other hand, the columns of $\tilde C$ and $\tilde C'$ generate the same lattice if and only if there exists a $3\times 3$ invertible matrix $Q$ such that
\begin{equation}
  \tilde CQ=\tilde C'\quad\iff\quad \tilde C=\tilde C'Q^{-1}
,\end{equation}
where both $Q$ and $Q^{-1}$ are integer matrices.
Such integer matrix $Q$ is called a \textit{unimodular matrix}, which has three equivalent definition: i) $Q\in\operatorname{GL}(3,\mathbb{Z})$, ii) $Q$ is invertible and $Q^{-1}$ is also an integer matrix, and iii) $\det Q=\pm 1$.
Therefore, $(M_A,M_B)$ and $(M_A',M_B')$ give the same $\tilde L$ if and only if $(M_A',M_B')=(M_AQ,M_BQ)$ where $Q$ is unimodular, in which case they also give the same $S$.
Since the total number of unimodular matrices is infinite, so is the supercell pair representations of $(S,\tilde L)$.
\subsection{IMT representation and HNF}\label{hnf}
In matrix theory, when performing column reduction on a real matrix, one uses a sequence of elementary column operations, which include: i) swapping two columns, ii) multiplying a column by a nonzero real number, and iii) adding a multiple of one column to another.
Consider a $3\times 3$ nonsingular real matrix, which can always be transformed into a lower triangular matrix with nonzero diagonal elements through i) and iii).
Then, each diagonal element can be reduced to $1$ via ii).
Finally, each lower off-diagonal element can be reduced to $0$ via iii), yielding an identity matrix.
The above process can be viewed as right-multiplying a sequence of elementary matrices.
Now we consider a \textit{restricted} version of elementary column operations: i') swapping two columns, ii') multiplying a column \textit{by} $\pm 1$, and iii') adding \textit{integer times} of a column to another.
Such operations on $3\times 3$ matrices can be viewed as right-multiplying a sequence of \textit{integer} elementary matrices whose inverses are also \textit{integer} matrices.
They generate all unimodular matrices by multiplication.
Using these restricted elementary column operations, a $3\times 3$ nonsingular \textit{integer} matrix $M$ can always be transformed into \textit{not} an identity matrix, but its Hermite normal form (HNF)
\begin{equation}
  H=\begin{bmatrix}
	  h_{11}&0&0\\h_{21}&h_{22}&0\\h_{31}&h_{32}&h_{33}
  \end{bmatrix},\quad 0\le h_{ij}<h_{ii}\pod{j<i}
.\end{equation}
In other words, we always have $M=HQ$ where $H$ is in HNF and $Q$ is unimodular.
Compared to the unrestricted column reduction, the diagonal elements in HNF are reduced only to positive numbers instead of $1$'s, which is a consequence of ii') being restricted.
The lower off-diagonal elements in HNF are reduced only to remainders instead of $0$'s, which is a consequence of iii') being restricted.
It is known that the HNF of an integer matrix is unique~\cite{cohen2013course}.
For nonsingular $M$, its unique HNF $H$ is always nonsingular, and thus $Q=H^{-1}M$ is also unique.
Therefore, the theorem used in the main text holds: any integer matrix $M$ with positive determinant can be uniquely decomposed into two integer matrices as $M=HQ$, where $H$ is in HNF and $\det Q=1$.
For a given $(S,\tilde L)$, denote one of its supercell pair representation by $(M_A,M_B)$.
By decomposing $M_\alpha=H_\alpha Q_\alpha$, we obtain the integer-matrix triplet (IMT) representation $(H_A, H_B, Q)$, where $H_\alpha$ is in HNF satisfying
\begin{equation}\label{deth}
	\det H_\alpha=\frac{\tilde Z}{Z_\alpha}\pod{\alpha=A\text{ or }B}
,\end{equation}
and $Q=Q_BQ_A^{-1}$ is unimodular.
As demonstrated in Subsection~\ref{unimodular}, any other supercell pair representation of $(S,\tilde L)$ can be written as $(M_AQ', M_BQ')$ where $Q'$ is unimodular.
Note that $M_\alpha Q'=H_\alpha (Q_\alpha Q')$, where $H_\alpha$ is in HNF and $Q_\alpha Q'$ is unimodular.
Therefore, its IMT representation is also $(H_A, H_B, (Q_BQ')(Q_AQ')^{-1})=(H_A,H_B,Q)$.
So far, we have proved that the IMT representation of $(S,\tilde L)$ is unique.
\subsection{Equivalence relation on SLMs}\label{equivslm}
For each $(M_A,M_B)$ satisfying Eq.~(\ref{detm}), there is an SLM given by Eq.~(\ref{scp}).
We denote the set of all SLMs with multiplicity $\mu$ by
\begin{align}
  \begin{split}
	  \mathbb{S}_\mu=\big\{&C_BM_BM_A^{-1}C_A^{-1}\,\big|\,M_\alpha\in\mathbb{Z}^{3\times 3},\\&\quad\det M_\alpha=\mu \operatorname{lcm}(Z_A,Z_B) / Z_\alpha\big\}
  .\end{split}
\end{align}
From Subsection~\ref{unimodular} we know that two choices of the primitive cell $C_A'$ and $C_A''$ differ in a right-multiplied unimodular matrix as $C_A''=C_A'Q$, since they generate the same $L_A$.
Therefore, the definition of $\mathbb{S}_\mu$ is independent of the choice of primitive cells.
Denote the space group of $\mathcal{A}$ and $\mathcal{B}$ by $G_A$ and $G_B$, respectively.
All pure rotations that have appeared in an element of $G_\alpha$ form a point group, which we denote by
\begin{equation}
	G_\alpha^\text{rot}=\big\{R\,\big|\,[R|\mathbf{t}]\in G_\alpha\big\}\pod{\alpha=A\text{ or }B}
.\end{equation}
Note that for each $R_\alpha\in G_\alpha^\text{rot}$, there is an alternative choice of primitive cell $R_\alpha C_\alpha$.
Hence, as long as $S\in\mathbb{S}_\mu$, we have $R_BSR_A^{-1}\in\mathbb{S}_\mu$.
We define the equivalence relation on $\mathbb{S}_\mu$ as
\begin{equation}\label{equiv}
  S'\sim S\quad\iff\quad\exists R_\alpha\in G_\alpha^\text{rot},\quad S'=R_BSR_A^{-1}
.\end{equation}
Whenever we find an SLM in $\mathbb{S}_\mu$, we immediately know its whole equivalence class.
The size of such equivalence class is
\begin{equation}\label{size}
	\big|\{R_BSR_A^{-1}\mid R_\alpha\in G_\alpha^\text{rot}\}\big|\le|G_A^\text{rot}|\times|G_B^\text{rot}|
,\end{equation}
and the SLM enumeration is accelerated by that many times.
A linear transformation $S$ with positive determinant has a singular value decomposition~\cite{axler1997linear}
\begin{equation}\label{svd}
  S=U\Sigma V^\text{T},\quad \Sigma=\begin{bmatrix}
	  s_1&0&0\\0&s_2&0\\0&0&s_3
  \end{bmatrix}
,\end{equation}
where $U,V\in\operatorname{SO}(3)$ are pure rotations and $s_1\ge s_2\ge s_3>0$.
It is also known that $S$ has a \textit{unique} polar decomposition~\cite{axler1997linear}
\begin{equation}\label{polar}
  S=R_SP_S
,\end{equation}
where $R_S=UV^\text{T}$ is a pure rotation, and $P_S=V\Sigma V^\text{T}=\sqrt{S^\text{T}S} $ is a positive-definite transformation scaling the system by $s_j\pod{j=1,2,3}$ along three orthogonal directions.
Physically, the PS is defined as $\varepsilon_j=s_j-1$.
From Eqs.~(\ref{equiv}) and (\ref{svd}) we see that equivalent SLMs have the same $s_j$, and hence the same PS.
They also yield the same orientation relationship (see Subsection~\ref{or}).
\subsection{Finiteness of SLMs with bounded strain}
Regardless of the criterion we use to bound the strain, e.g., the pricipal strain $\varepsilon_j$, or the root-mean-square strain (RMSS), we can always consider a looser bound of the form
\begin{equation}
	\sigma_1(S)\le s_\text{max}
,\end{equation}
where $\sigma_j(S)$ is the $j$-th largest singular value of $S$, and $s_\text{max}$ is a finite positive number.
Applying the inequality $\sigma_1(XY)\le\sigma_1(X)\sigma_1(Y)$ to
\begin{equation}
  Q=H_B^{-1}C_B^{-1}SC_AH_A
,\end{equation}
we have
\begin{align}
  \begin{split}
	  \sigma_1(Q)&\le \sigma_1(H_B^{-1})\sigma_1(C_B^{-1})\sigma_1(C_A)\sigma_1(H_A)\sigma_1(S)\\
		     &\le\frac{\sigma_1(C_A)\sigma_1(H_A)}{\sigma_3(C_B)\sigma_3(H_B)}\,s_\text{max}
  .\end{split}
\end{align}
Under fixed $H_A$ and $H_B$, we have $\sigma_1(Q)$ bounded.
Noticing that the absolute value of a matrix element of $Q$ is never larger than $\sigma_1(Q)$, the number of possible $Q$'s is finite.
On the other hand, given the multiplicity $\mu$ or $\tilde Z$, the number of $(H_A,H_B)$'s satisfying Eq.~(\ref{deth}) is finite, as the total number of matrices in HNF with determinant $m$ is~\cite{hart2008algorithm}
\begin{equation}\label{nhnf}
	n_\text{HNF}(m)=\sum_{k\in D_m}k\cdot|D_k|
,\end{equation}
where $D_m=\{1,\cdots,m\}$ is the set of all divisors of $m$.
Since each SLM is associated with a unique $(H_A,H_B,Q)$, the size of $\{\sigma_1(S)<s_\text{max}\mid S\in\mathbb{S}_\mu\}$ is finite.
\section{Geometric criteria}\label{geo}
To quantify how likely a CSM is to occur in reality, the ideal way is to take full energetics into account.
However, geometric criteria are computationally cheaper~\cite{sadeghi2013metrics,bartok2013representing,ferre2015permutation}, and can be used to filter plausible CSMs from the vast candidates efficiently.
Previous study indicates that SSPT favors the CSM that minimizes \textit{the total distance traveled by the atoms}~\cite{therrien2020minimization}.
We separately consider the contributions of i) the lattice deformation determined by $S$, and ii) the periodic atom displacement in $\mathcal{J}^S$.
These two parts scale differently as the system size $N$ increases: for the root-mean-square (Euclidean) distance
\begin{equation}\label{d0}
	d=\sqrt{\sum_{j=1}^N |\mathbf{r}_j'-\mathbf{r}_j|^2} 
,\end{equation}
where $\mathbf{r}_j$ and $\mathbf{r}_j'$ are the initial and final position of the $j$-th atom, Therrien \textit{et al.}~have derived that~\cite{therrien2020matching}
\begin{equation}
	d^2=gN^{\frac{5}{3}}+hN
,\end{equation}
where $g$ is alone determined by i), while $h$ is contributed by both i) and ii).
However, their summations over $j$ are done within specified cells, making $g$ and $h$ dependent on the supercell selection.
In this section, we consider a ball-shaped summation region to obtain isotropic versions of $\bar g$ and $\bar h$, which are independent of the cell selection and thus suitable for evaluating CSMs.
To score i) the SLM $S$ and ii) the periodic $\mathcal{J}^S$ separately, we derive their contributions to Eq.~(\ref{d0}) when they each exist alone.
We emphasize that given the initial structure $\mathcal{A}$, the final structure produced by SSPT may not be strictly $\mathcal{B}$, but can also be $R\mathcal{B}+\boldsymbol{\tau}$, where $R\in\operatorname{SO}(3)$ and $\boldsymbol{\tau}\in\mathbb{R}^3$ are arbitrary (see Subsection~\ref{or}).
To quantify the likelihood of a CSM $\mathcal{J}$, one should always consider the ``best'' way for $\mathcal{J}$ to occur.
In other words, rather than the root-mean-square of $|\mathcal{J}(\mathbf{r}_j)-\mathbf{r}_j|$, the geometric criterion should be
\begin{equation}\label{d}
	d=\min_{R,\boldsymbol{\tau}}\sqrt{\sum_{j=1}^N |R\mathcal{J}(\mathbf{r}_j)+\boldsymbol{\tau}-\mathbf{r}_j|^2}
.\end{equation}
In such definition, Eq.~(\ref{d}) no longer depends on the artificially specified position and orientation of $\mathcal{B}$, nor does it depend on those of $\mathcal{A}$.
So far, comparing the CSMs between fixed $\mathcal{A}$ and $\mathcal{B}$ is sufficient to discuss the various CSMs between $\mathcal{A}$ and $R\mathcal{B}+\boldsymbol{\tau}$ established by SSPT.
\subsection{RMSS: geometric criterion for SLM}\label{criteria1}
The linear transformation $S$ itself can be viewed as a CSM between $\mathcal{A}$ and $S\mathcal{A}$.
To derive the coefficient
\begin{equation}
	\bar g=\lim_{N\to\infty}\frac{d^2}{N^{\frac{5}{3}}}
,\end{equation}
we consider
\begin{equation}\label{dg}
	d_{g}=\min_R\sqrt{\sum_{j=1}^N|RS\mathbf{r}_j-\mathbf{r}_j|^2}
,\end{equation}
where the contribution of $\boldsymbol{\tau}$ is omitted since it is only proportional to $N$.
Note that
\begin{align}\label{srr}
  \begin{split}
	  |RS\mathbf{r}_j-\mathbf{r}_j|^2
	  &=(RS\mathbf{r}_j-\mathbf{r}_j)^\text{T}(RS\mathbf{r}_j-\mathbf{r}_j)
	\\&=\mathbf{r}_j^\text{T}(RS-I)^\text{T}(RS-I)\mathbf{r}_j
	\\&=\lambda_1x_j^2+\lambda_2y_j^2+\lambda_3z_j^2
,\end{split}
\end{align}
where $\lambda_1,\lambda_2,\lambda_3>0$ are eigenvalues of $(RS-I)^\text{T}(RS-I)$, and $(x_j,y_j,z_j)$ are coordinates of $\mathbf{r}_j$ under an orthonormal eigenbasis.
We have
\begin{equation}\label{df}
	\sum_{j=1}^N|RS\mathbf{r}_j-\mathbf{r}_j|^2=N\left(\lambda_1\overline{x^2}+\lambda_2\overline{y^2}+\lambda_3\overline{z^2}\right),
\end{equation}
where
\begin{equation}
	\overline{x^2}=\frac{1}{N}\sum_{j=1}^Nx_j^2
.\end{equation}
Considering a ball region with radius $\zeta\to\infty$, the distributions of $x_j /\zeta$, $y_j /\zeta$, and $z_j /\zeta$ converge to
\begin{equation}
	f(u)=\frac{3}{4}(1-u^2),\quad -1\le u\le 1
,\end{equation}
so that
\begin{equation}
	\frac{1}{N}\sum_{j=1}^N\left(\frac{x_j}{\zeta}\right)^2\to\int_{-1}^1 u^2f(u)\,\mathrm{d}u=\frac{1}{5}
.\end{equation}
Using $N\sim\rho_A\cdot \frac{4}{3}\pi \zeta^3$, Eq.~(\ref{df}) becomes
\begin{align}
  \begin{split}
	&\quad\,\sum_{j=1}^N|RS\mathbf{r}_j-\mathbf{r}_j|^2
	\\&\sim \frac{1}{5}\left(\frac{3}{4\pi \rho_A}\right)^{\frac{2}{3}}\left(\lambda_1+\lambda_2+\lambda_3\right)N^{\frac{5}{3}}
      \\&=\frac{1}{5}\left(\frac{3}{4\pi \rho_A}\right)^{\frac{2}{3}}\operatorname{Tr}\!\left[(RS-I)^\text{T}(RS-I)\right]N^{\frac{5}{3}}
  .\end{split}
\end{align}
Therefore, Eq.~(\ref{dg}) is determined by the $R$ that minimizes the trace
\begin{align}
  \begin{split}
	&\quad\,\operatorname{Tr}\!\left[(RS-I)^\text{T}(RS-I)\right]
	\\&=\operatorname{Tr}\!\left[S^\text{T}S+I-RS-(RS)^\text{T}\right]
	\\&=s_1^2+s_2^2+s_3^2+3-2\operatorname{Tr}(RS)
  .\end{split}
\end{align}
Von Neumann's trace inequality states that~\cite{mirsky1975trace}
\begin{equation}
	\operatorname{Tr}(RS)\le\sum_{k=1}^3\sigma_k(R)\sigma_k(S)=s_1+s_2+s_3
,\end{equation}
where equality holds if and only if $R=R_S$~\cite{carlsson2021neumann}, which is defined in Eq.~(\ref{fs}).
In other words, the ``best'' final structure is $P_S\mathcal{A}=\sqrt{S^\text{T}S}\,\mathcal{A}$, in which case we have $\lambda_k=\varepsilon_k^2$ and
\begin{equation}
	d_g^2\sim\frac{1}{5}\left(\frac{3}{4\pi\rho_A}\right)^{\frac{2}{3}}\left(\varepsilon_1^2+\varepsilon_2^2+\varepsilon_3^2\right) N^{\frac{5}{3}}
.\end{equation}
Thus, we obtain
\begin{equation}
	\bar g=\frac{1}{5}\left(\frac{3}{4\pi\rho_A}\right)^{\frac{2}{3}}\left(\varepsilon_1^2+\varepsilon_2^2+\varepsilon_3^2\right)
.\end{equation}
The root-mean-square strain (RMSS) is defined as
\begin{equation}
	\mathrm{RMSS}=\sqrt{\frac{\varepsilon_1^2+\varepsilon_2^2+\varepsilon_3^2}3}
,\end{equation}
which is proportional to $\sqrt{\bar g}$ and $d_g$.
\subsection{RMSD: geometric criterion for periodic CSM}\label{criteria2}
For the induced CSM $\mathcal{J}^S\colon S\mathcal{A}\to\mathcal{B}$ with translation group $\tilde T$, we say two atoms in $S\mathcal{A}$ to be equivalent if they do not coincide under any translation in $\tilde T$.
We consider the total distance traveled by $\tilde Z$ inequivalent atoms (e.g., the atoms in a common supercell $\tilde C=S\tilde C_A=\tilde C_B$).
Since $\mathcal{J}^S$ is already periodic without further deformation, the SLM of $\mathcal{J}^S$ is $I$.
According to Eq.~(\ref{fs}), the rotation-free orientation that minimizes the total distance traveled by  is $R_I=I$.
We therefore let $R=I$ in Eq.~(\ref{df}), obtaining
\begin{equation}\label{dh}
d_h=\min_{\boldsymbol{\tau}}\sqrt{\sum_{j=1}^{\tilde Z}|\mathbf{r}_j'+\boldsymbol{\tau}-\mathbf{r}_j|^2}
,\end{equation}
where $\mathbf{r}_j\in S\mathcal{A}$ and $\mathbf{r}_j'=\mathcal{J}^S(\mathbf{r}_j)$.
Denoting the average displacement of $\mathbf{r}_j'$ from $\mathbf{r}_j$ by
\begin{equation}
	\boldsymbol{\tau}_0=\frac{1}{\tilde Z}\sum_{j=1}^{\tilde Z}(\mathbf{r}'_j-\mathbf{r}_j)
,\end{equation}
we have
\begin{widetext}
\begin{align}
  \begin{split}
	  \sum_{j=1}^{\tilde Z}|\mathbf{r}_j'+\boldsymbol{\tau}-\mathbf{r}_j|^2
	  &=\sum_{j=1}^{\tilde Z}\left[(\boldsymbol{\tau}+\boldsymbol{\tau}_0)+(\mathbf{r}_j'-\mathbf{r}_j-\boldsymbol{\tau}_0)\right]^2
	\\&=\tilde Z(\boldsymbol{\tau}+\boldsymbol{\tau}_0)^2+2(\boldsymbol{\tau}+\boldsymbol{\tau}_0)\cdot\sum_{j=1}^{\tilde Z}(\mathbf{r}_j'-\mathbf{r}_j-\boldsymbol{\tau}_0)+\sum_{j=1}^{\tilde Z}(\mathbf{r}_j'-\mathbf{r}_j-\boldsymbol{\tau}_0)^2
	\\&=\tilde Z(\boldsymbol{\tau}+\boldsymbol{\tau}_0)^2+\sum_{j=1}^{\tilde Z}(\mathbf{r}_j'-\mathbf{r}_j-\boldsymbol{\tau}_0)^2
 ,\end{split}
\end{align}
\end{widetext}
which is minimized when $\boldsymbol{\tau}=-\boldsymbol{\tau}_0$, i.e., when the average displacement is canceled out.
Since atoms equivalent with $\mathbf{r}_j$ satisfy Eq.~(\ref{periodj}), we have
\begin{equation}
	\forall\mathbf{t}\in\tilde T,\quad \mathcal{J}^S(\mathbf{r}_j+\mathbf{t})-(\mathbf{r}_j+\mathbf{t})=\mathbf{r}_j'-\mathbf{r}_j
.\end{equation}
Therefore, $d_h^2=\bar h N$ holds as long as the summation is done over \textit{whole sets} of inequivalent atoms, where
\begin{equation}
	\bar h=\frac{1}{\tilde Z}\sum_{j=1}^{\tilde Z}|\mathbf{r}_j'-\boldsymbol{\tau}_0-\mathbf{r}_j|^2 
.\end{equation}
The root-mean-square displacement (RMSD) is defined as $\sqrt{\bar h}$, which is proportional to $d_h$.